\documentclass[useAMS,usenatbib]{mn2e}
\usepackage{amsmath}
\usepackage{txfonts}
\usepackage{graphics}
\usepackage{graphicx}
\usepackage{times}
\usepackage{placeins}
\usepackage{lastpage}

\newcommand{\mnras}{MNRAS}

\newcommand{\Mpcoh}{{\rm Mpc}\ h^{-1}}

\begin{document}

\title[Deviations from $\Lambda$CDM and GR]{The Clustering of Galaxies in the
SDSS-III DR9 Baryon Oscillation Spectroscopic Survey: Testing Deviations from
$\Lambda$ and General Relativity using anisotropic clustering of galaxies}

\author[Samushia et al.]{
\parbox{\textwidth}{
Lado Samushia$^{1,2}$\thanks{E-mail: lado.samushia@port.ac.uk}, 
Beth A. Reid$^{3,4}$, 
Martin White$^{3,5}$, 
Will J. Percival$^{1}$,
Antonio J. Cuesta$^{6}$,
Lucas Lombriser$^{1}$,
Marc Manera$^{1}$,
Robert C. Nichol$^{1,7}$,
Donald P. Schneider$^{8,9}$,
Dmitry Bizyaev$^{10}$,
Howard Brewington$^{10}$,
Elena  Malanushenko$^{10}$, 
Viktor Malanushenko$^{10}$,
Daniel Oravetz$^{10}$,
Kaike Pan$^{10}$, 
Audrey Simmons$^{10}$, 
Alaina Shelden$^{10}$,
Stephanie Snedden$^{10}$,
Jeremy L. Tinker$^{11}$,
Benjamin A. Weaver$^{11}$,
Donald G. York$^{12}$,
Gong-Bo Zhao$^{1,13}$.
} \\
\vspace*{4pt} \\
$^{1}$ Institute of Cosmology and Gravitation, University of Portsmouth, Dennis Sciama Building, Portsmouth, P01 3FX, U.K.  \\
$^{2}$ National Abastumani Astrophysical Observatory, Ilia State University, 2A Kazbegi Ave., GE-1060 Tbilisi, Georgia  \\
$^{3}$ Lawrence Berkeley National Laboratory, 1 Cyclotron Road, Berkeley, CA 94720, USA \\
$^{4}$ Hubble Fellow \\
$^{5}$ Departments of Physics and Astronomy, University of California, Berkeley, CA 94720, USA \\
$^{6}$ Department of Physics, Yale University, 260 Whitney Ave, New Heaven, CT 06520, USA \\
$^{7}$ SEPNet \\
$^{8}$ Department of Astronomy and Astrophysics, The Pennsylvania State University, University Park, PA 16802, USA \\
$^{9}$ Institute for Gravitation and the Cosmos, The Pennsylvania State University, University Park, PA 16802, USA \\
$^{10}$ Apache Point Observatory, P.O. Box 59, Sunspot, NM 88349-0059, USA \\
$^{11}$ Center for Cosmology and Particle Physics, New York University, New York, NY 10003, USA \\
$^{12}$ Department of Astronomy and Astrophysics and Enrico Fermi Institute, 5640 So. Ellis Ave, Chicago, IL 60615, USA \\
$^{13}$ National Astronomy Observatories, Chinese Academy of Science, Beijing, 100012, P.R.China \\
}

\date{\today} 
\pagerange{\pageref{firstpage}--\pageref{lastpage}}

\maketitle

\label{firstpage}

\begin{abstract}
We use the joint measurement of geometry and growth from anisotropic galaxy
clustering in the Baryon Oscillation Spectroscopic Survey (BOSS) Data Release 9
(DR9) CMASS sample reported by \citeauthor{Reid12} to constrain dark energy (DE)
properties and possible deviations from the General Relativity (GR).  Assuming
GR and taking a prior on the linear matter power spectrum at high redshift from
the cosmic microwave background (CMB), anisotropic clustering of the CMASS DR9
galaxies alone constrains $\Omega_{\rm m} = 0.308 \pm 0.022$ and $100\Omega_{\rm
k} = 5.9 \pm 4.8$ for $w = -1$, or $w = -0.91 \pm 0.12$ for $\Omega_k = 0$.
When combined with the full CMB likelihood, the addition of the anisotropic
clustering measurements to the spherically-averaged BAO location increases the
constraining power on dark energy by a factor of 4 in a flat CDM cosmology with
constant dark energy equation of state $w$ (giving $w = -0.87 \pm 0.05$).  This
impressive gain depends on our measurement of both the growth of structure and
Alcock-Paczynski effect, and is not realised when marginalising over the
amplitude of redshift space distortions.  Combining with both the CMB and
Supernovae Type Ia (SNeIa), we find $\Omega_{\rm m} = 0.281 \pm 0.014$ and
$1000\Omega_{\rm k}=-9.2\pm5.0$ for $w = -1$, or $w_0 = -1.13 \pm 0.12$ and
$w_{\rm a}=0.65 \pm 0.36$ assuming $\Omega_k = 0$.   Finally, when a
$\Lambda$CDM background expansion is assumed, the combination of our estimate of
the growth rate with previous growth measurements provides tight constraints on
the parameters describing possible deviations from GR giving $\gamma = 0.64 \pm
0.05$.  For one parameter extensions of the flat $\Lambda$CDM model, we find a
$\sim 2\sigma$ preference either for $w > -1$ or slower growth than in GR.
However, the data is fully consistent with the concordance model, and the
evidence for these additional parameters is weaker than $2\sigma$.
\end{abstract}

\section{Introduction}

All currently available cosmological observations, including the latest datasets
of CMB temperature and polarisation anisotropies \citep{Komatsu2011}, Supernovae
Type Ia (SNeIa) magnitudes \citep{Suzuki2011} and the distance ladder mapped
by Baryon Acoustic Oscillation (BAO) peak signature in the clustering of
galaxies \citep{Aadvark:2012} are consistent with a simple cosmological model in
which general relativity (GR) describes gravitational interactions on all
scales and times, about 70 per cent of the Universe's current energy density is in form of
a Cosmological Constant as originally described by Einstein, and most of the
remaining 30 per cent is in form of nonrelativistic ``dark matter'' \citep[For a
detailed review see e.g.,][]{Peebles2003,Weinberg2012}.

Ongoing and future observations have been designed to test if the cosmological
constant needs to be replaced by a dynamical Dark Energy (DE), and if so, to
measure the properties of this DE fluid. We should also be able to tell if GR
describes the properties of gravity on cosmological scales or if it must be
replaced by a yet unknown modified theory of gravity (MG) \citep[see
e.g.,][]{Albrecht2009,Zhao2009,Samushia11,Wang2010}.  Observational effects of
dynamic DE and MG are partially degenerate and careful data analysis should take
into account both possibilities \citep{Ishak:2006,Shapiro:2010}.

The clustering of galaxies provides a very powerful and robust test of both the
nature of DE and MG. The shape of the measured
correlation function \citep{Rei10} or the power-spectrum \citep{Montesano2011},
analogously to the shape of the measured CMB power-spectrum \citep{WMAP7}, can
be used to constrain basic cosmological parameters. Features within
the clustering signal, particularly the BAO, allow the
clustering to be used as a standard ruler. Additionally, although the
statistical properties of galaxy clustering are expected to be isotropic,
the measured clustering can be highly anisotropic, depending on how
redshifts are translated to distances. The two main sources of
this apparent anisotropy are redshift-space distortions
\citep[RSD;][]{Kai87} and the Alcock-Paczynski \citep[AP;][]{AP} effect.

RSD arise because peculiar velocities contribute to observed galaxies redshifts,
but can not be corrected for when computing line-of-sight separations. On quasi-linear
scales, the average pairwise galaxy peculiar velocity is negative, meaning that
galaxies are on average falling towards the mass over-densities traced by
neighbouring galaxies. These coherent motions appear as a ``squashing'' of the
correlation function along the line-of-sight. The amplitude of the observed
anisotropy can be used to infer the strength of the gravitational interaction at
different scales and redshifts. \citep[For a detailed review of RSD
see][]{HamiltonReview}. RSD allow measurements of the amplitude of fluctuations
in the velocity field, which in linear theory give a dependence on 
\begin{equation} 
  f(z)\sigma_8(z)=\frac{d\sigma_8(z)}{d\ln a}, 
  \label{eq:fs8} 
\end{equation} 
\noindent 
where $\sigma_8(z)$ is the overall normalisation of the matter density
fluctuations.

The AP effect results from the fact that, to convert observed angular positions
and redshifts of galaxies into physical positions, we must use a cosmological
model on the observed lightcone. If the wrong model is used when computing the
correlation function, the initial isotropy of the clustering signal will be
distorted. The measured anisotropy of clustering can be used to infer the proper
geometry and hence the true values of cosmological parameters.
If we have a prior on the shape of the correlation function, the dilation of
scales between the spherically averaged observed and model clustering statistics
allows a measurement of
\begin{equation}
  D_V(z) = \left[(1+z)^2D_A^2(z)\frac{cz}{H(z)}\right]^{1/3},
  \label{eq:Dv}
\end{equation}
\noindent
where $D_{\rm A}(z)$ is the physical angular diameter distance and $H(z)$ is the
Hubble expansion rate \citep{Eis05}.  Applying the AP test to the measured
direction dependent clustering additionally allows the measurement of
\begin{equation}
  F(z) = \frac{1+z}{c}D_{\rm A}(z)H(z).
  \label{eq:F}
\end{equation}
\noindent
This allows the degeneracy between radial and angular distances in BAO
measurements to be broken and an accurate estimate of Hubble expansion rate at
different redshifts to be derived.

Many RSD measurements have been made from a variety of galaxy surveys, including
most recently the 2dFGRS \citep{Per04}, the VVDS \citep{Guz08}, the 2SLAQ
\citep{Ang08}, the SDSS-II \citep{Cabre:2009,Song:2011,SamPerRac11}, the
WiggleZ \citep{Bla11a}, and the 6dFGRS \citep{Beutler12} surveys.
These measurements have in turn been used to set constraints on the
cosmological growth rate.

Thus far geometric constraints from galaxy clustering have focused predominantly
on spherically averaged power spectra or correlation functions. From such
measurements, the BAO feature allows few percent-level distance measurements
\citep[for the latest constraints,
see][]{Per10,Bla11b,Beutler11,Padmanabhan2012,Aadvark:2012}.  By contrast, the
AP test has received less attention, since better signal-to-noise data is needed
to disentangle RSD and AP effects, and more careful modelling of anisotropic
correlation function is required. The AP effect has been recently used to
jointly measure $D_A$, $H$ and $f\sigma_8$  in three redshift bins from WiggleZ
survey \citep{Blake2012} and using the SDSS-II LRG sample \citep{Chuang2012}. 

The Sloan Digital Sky Survey \citep[SDSS;][]{York:2000} has mapped over one
third of the sky using the dedicated 2.5-m Sloan telescope \citep{Gunn:2006}. A
drift-scanning mosaic CCD camera \citep{Gunn:1998} imaged sky in five
photometric bandpasses \citep{Fukugita:1996} to a limiting magnitude of $r
\simeq 22.5$.  The ongoing Baryon Oscillation Spectroscopic Survey
\citep[BOSS;][]{BOSS}, which is a part of SDSS-III \citep{Eis11}, is measuring
spectra of 1.5 million galaxies and 150,000 quasars selected from the multicolor
SDSS imaging. The CMASS sample in the BOSS Data Release 9 (DR9) contains a
catalog of $264\,283$ highly biased ($b \sim 2$) galaxies sampling an effective
volume of about $0.77\ h^{-3}\ {\rm Gpc}^3$ with mean redshift of $z\sim 0.57$,
allowing for best-to-date statistical uncertainty in the measurement of galaxy
clustering \citep[see][for detailed description of early data]{Whi11}.

This work is part of series of papers providing a comprehensive description of
the galaxy clustering in the CMASS DR9 sample. \citet{Nuza:2012} compared
the clustering of CMASS DR9 galaxies to state of the art dark matter simulations and
showed that they are compatible with $\Lambda$CDM model and GR.
\citet{Ross:2011} provided an up-to-date description of the CMASS DR9 data, a study
of possible observational systematic effects and the methods to remove known
systematics.  \citet{Manera:2012a} presented 600 mock catalogs that match the
observed volume of CMASS DR9 sample and are essential for determining the covariance
matrix associated with the measured correlation functions and power
spectra.  \citet{Aadvark:2012} used these data to measure the BAO peak
position to 1.7 per cent precision.  \citet{Sanchez:2012} derived cosmological
constraints using the full shape of the measured spherically-averaged
correlation function, while \citet{Reid12} studied the anisotropic clustering of
CMASS DR9 galaxies using the measured monopole and quadrupole moments of
the correlation function (henceforth we shall simply refer to these as
the monopole and quadrupole). 

By adopting a sophisticated model for galaxy clustering in the quasi-linear
regime, \citet{Reid12} made accurate RSD and AP measurements from the
direction-dependent clustering of CMASS DR9 galaxies to simultaneously measure
growth and geometry at a redshift of $z = 0.57$. We now extend this work to
investigate the cosmological implications of these measurements. We show that
information provided by the RSD-derived growth rate significantly enhances
constraints on basic cosmological parameters compared to the case where only
geometric measurements are used. We combine measurements of the growth rate,
angular diameter distance and expansion rate with previous measurements to
constrain properties of DE and gravity. In combination with CMB, $H_0$ and SNeIa
data we are able to estimate values of basic cosmological parameters to very
high precision and tightly constrain possible deviations from $\Lambda$CDM and
GR (henceforth we will refer to the model in which the background expansion
follows $\Lambda$CDM and the gravity is described by GR as $\Lambda$CDMGR).

The paper is organised as follows: in Sec.~\ref{sec:measurements} we
describe the CMASS DR9 AP and RSD measurements, in
Sec.~\ref{sec:previous} we briefly review previous similar
measurements, in Sec.~\ref{sec:theory} we describe different ways of
looking for deviations from $\Lambda$CDMGR, and in
Sec.~\ref{sec:constraints} we present constraints on deviations form
$\Lambda$CDMGR. We conclude and discuss our results in
Sec.~\ref{sec:conclusion}.

\section{CMASS DR9 Measurements}
\label{sec:measurements}

\citet{Reid12} used the measured anisotropic correlation function of galaxies in
the CMASS DR9 sample to jointly estimate $D_{\rm V}(z=0.57)$, $F(z=0.57)$ and
$f\sigma_8(z=0.57)$. To obtain these estimates they fitted the monopole and
quadrupole in the range of scales $25 \Mpcoh < s < 160 \Mpcoh$. The theoretical
model correlation function was computed from the ``streaming model'' of
\citet{ReiWhi11}. Linear theory predictions for the anisotropic correlation
function cease to be accurate on relatively large scales \citep[for the
importance of nonlinear corrections see and different ways of modeling RSD in
quasi-nonlinear regime see,
e.g.,][]{delaTorre2012,Matsubara2011,Taruya2010,Taruya2011,Okamura2011}.
\citet{ReiWhi11} demonstrated that the ``streaming model'' used as the
theoretical basis of our CMASS DR9 analysis reproduces the monopole and
quadrupole measured in Nbody simulations with better than 2 per cent precision
down to the scales of $25 \Mpcoh$; better than the level demanded by DR9 data.
Appendices B3 and B4 of \citet{Reid12} further quantify the (small) theoretical
errors, including uncertainties in the ``Fingers-of-God'', using realistic mock
galaxy catalogs.  The shape of the linear real-space correlation function was
marginalised over using a WMAP7 prior on parameters $\Omega_{\rm m}h^2$,
$\Omega_{\rm b}h^2$ and $n_{\rm s}$. This triplet of parameters fixes relative
clustering at different physical separations; the geometrical measurements of
$D_{\rm V}$ and $F$ result from matching the measured shape of the galaxy
correlation function to the shape fixed by CMB data while most of information
about $f\sigma_8$ comes from the relative amplitudes of the measured monopole
and quadrupole.  \citet{Reid12} considered a number of models with different
assumptions about how the shape of the correlation function relates to the
background geometry and growth. Most relevant to this paper is the model in
which the shape of the correlation function is fixed by a WMAP7 prior but the
expansion of the Universe at lower redshifts and the growth of structure since
recombination are allowed to be arbitrary (Model $4$ in \citealt{Reid12}). Joint
estimates of $f\sigma_8$, $D_{\rm A}$ and $H$ recovered by applying this model
are independent of assumptions about either the late-time behaviour of DE or the
structure growth and can be used to constrain models with arbitrary DE Equation
of State (EoS) and growth history as long as physics at recombination is
unaltered and linear growth remains scale-independent. The \citet{Reid12}
measurements have already been used to constrain Galileon gravity
\citep{Appleby2012} and cosmic growth index \citep{Hudson2012,Rapetti2012}.

The most general way of testing DE and MG models against anisotropic clustering
measurements made with the CMASS sample would be to fit predictions of the
models to the measurements of monopole and quadrupole directly. We have checked
that in all relevant cases simply assuming a 3-dimensional multi-variable
Gaussian likelihood around the $f\sigma_8$, $D_{\rm V}$ and $F$ measurements of
Model $4$ retains most of the constraining power of the data set and results in
very similar constraints compared with exploring the full likelihood. In fact,
while the likelihood of the measurements of \citet{Reid12} is close to a
multivariate Gaussian, it is slightly skewed towards higher values of $F$ and
$f\sigma_8$. We have checked that the effect of the skewness on the derived
likelihood of cosmological parameters is small and only affects regions of low
likelihood (Appendix~\ref{appendixA}).  For the rest of the paper we will ignore this small
effect. 

Although the measurements of geometry and growth rate in
\citet{Reid12} were derived using a WMAP ``shape'' prior, the
resulting estimates are not correlated with the WMAP parameters
$\Omega_{\rm m}h^2$, $\Omega_{\rm b}h^2$ and $n_{\rm s}$. This allows
us to combine these measurements with the WMAP likelihood without
double-counting the CMB information. We have checked that combining
Model $4$ measurements with WMAP data gives very similar likelihood
surfaces as those calculated by jointly fitting the CMASS DR9 monopole
and quadrupole and the CMB data. This approximation allows us to
greatly speed up the calculation of likelihoods for a variety of
cosmological models.

\section{Previous measurements of growth and geometry from anisotropic
clustering}
\label{sec:previous}

\begin{figure}
  \includegraphics[width=90mm]{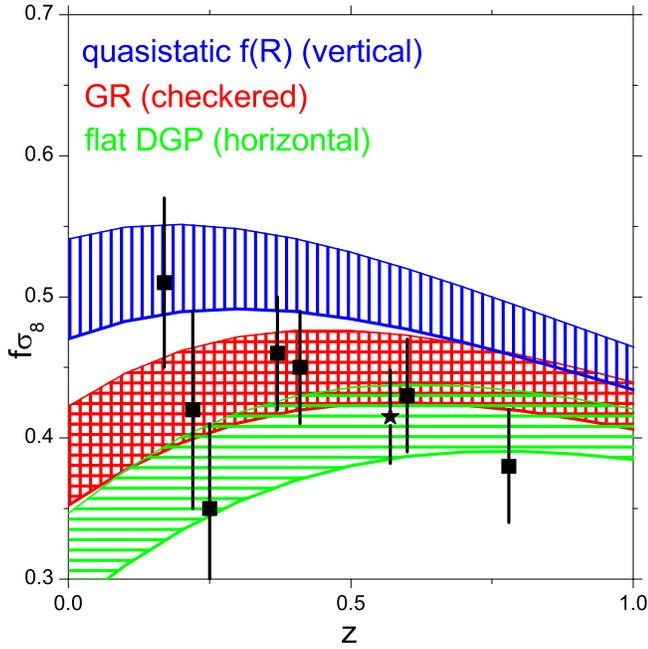}
  \caption{The data points show the CMASS DR9 measurement of $f\sigma_8$
  (circle) along with similar, low redshift, measurements (squares) and
  $1\sigma$ errorbars as presented in Table~\ref{table:fs8constraints}. The three stripes
  show theoretical predictions for different gravity models allowing
  for uncertainty in the background cosmological parameters,
  constrained using only the WMAP 7 data \citep{Komatsu2011}.}
  \label{fig:fs8}
\end{figure}

The growth rate of structure has now been measured from the RSD
signal in the galaxy clustering pattern, at a range of redshifts from
$z\sim0.1$ up to $z\sim0.7$.  Fig.~\ref{fig:fs8} shows the most recent
CMASS DR9 estimates of $f\sigma_8$ alongside previous measurements
from other surveys that are used later on in our
analysis. Measurements, $1\sigma$ errorbars and references to original
publications are presented in Table~\ref{table:fs8constraints}.

\begin{table}
  \begin{center}
    \begin{tabular}{llll}
      z & $f\sigma_8$ & survey & reference\\
        \hline
	0.17 & $0.51 \pm 0.06$ & 2dFGRS & \protect\citet{Per04}\\
	0.22 & $ 0.42 \pm 0.07 $ & WiggleZ & \protect\citet{Bla11a}\\
	0.25 & $0.35 \pm 0.06$ & SDSS LRG & \protect\citet{SamPerRac11}\\
	0.37 & $0.46 \pm 0.04$ & SDSS LRG & \protect\citet{SamPerRac11}\\
	0.41 & $ 0.45 \pm 0.04 $ & WiggleZ & \protect\citet{Bla11a}\\
	0.57 & $0.43 \pm 0.03$ & BOSS CMASS & \protect\citet{Reid12}\\
	0.6 & $ 0.43 \pm 0.04$ & WiggleZ & \protect\citet{Bla11a}\\
	0.78 & $ 0.38 \pm 0.04$ & WiggleZ & \protect\citet{Bla11a}\\
        \hline
        0.067 & $0.42 \pm 0.06$ & 6dFGRS & \protect\citet{Beutler12}\\
        0.77 & $0.49 \pm 0.18$ & VVDS & \protect\citet{Guz08}\\
      \end{tabular}
    \end{center}
    \caption{Compilation of RSD based $f\sigma_8$ measurements. Note that the
      6dFGRS measurement \protect\citet{Beutler12} was too recent to be
      included in our analysis, while the measurement of
      \protect\citet{Guz08} was excluded due to more recent, stronger, measurements
      at similar redshifts.}
    \label{table:fs8constraints}
\end{table}

The bands on Fig.~\ref{fig:fs8} show theoretical predictions of GR, DGP
\citep{Dvali2010} and $f(R)$ \citep{Buchdahl1970} theories of gravity when a
WMAP7 prior is applied to cosmological parameters describing expansion of the
Universe. For the DGP predictions we ignore the scale dependence of growth rate on
large scales while for the $f(R)$ we use the model studied in
\citet{Starobinsky2007}; the general $f(R)$ model has a clear GR limit and by
tuning model parameters its predictions can be made arbitrarily close to the GR
predictions.

\begin{figure*}
  \includegraphics[width=180mm]{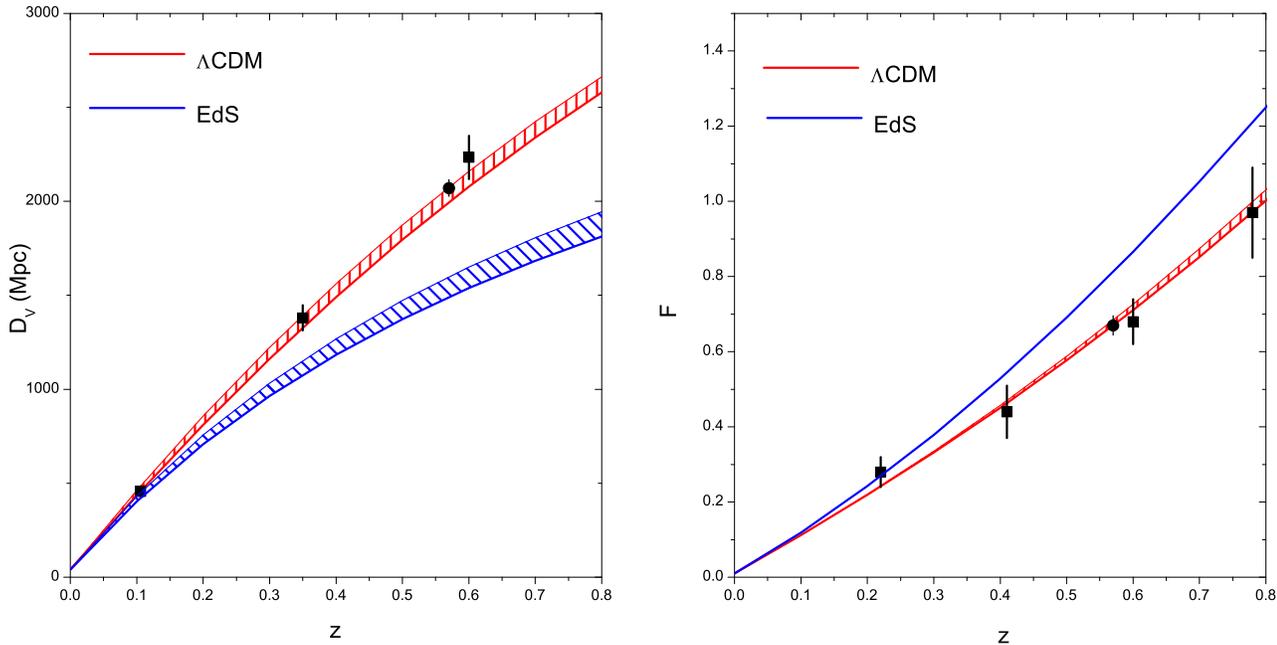}
  \caption{The data points show recent measurements of $D_V$ (left panel,
  squares) and $F$ (right panel, squares) described in Table 1 along with latest
  measurements from CMASS DR9 data (circle) with $1\sigma$ errorbars. Red
  stripes show theoretical prediction of spatially-flat $\Lambda$CDM within the
  uncertainty in basic cosmological parameters as measured by WMAP 7 data; Blue
  stripes show theoretical predictions of Einstein-deSitter model within
  the uncertainty in $H_0$ as measured by HST key project \citep{riess:2011}.}
  \label{fig:dvfm}
\end{figure*}

Geometrical measurements are also available from the local Universe up to
redshift of $z\sim 0.8$. Fig.~\ref{fig:dvfm} shows the most recent
CMASS DR9
estimates of $D_V$ and $F$ along with similar results from other
surveys. The numerical values of these estimates, $1\sigma$ errorbars and
references to original publications are presented in Tables
\ref{table:Dvconstraints} and \ref{table:Hconstraints}. 

\begin{table}
  \begin{center}
    \begin{tabular}{llll}
      {\it z} & $D_V\ {\rm Mpc}^3$ & survey & reference\\
        \hline
	0.106 & $456 \pm 27$ & 6dFGRS & \protect\citet{Beutler11}\\
	0.35 & $1380 \pm 68$ & SDSS LRG & \protect\citet{Rei10}\\
	0.57 & $2071 \pm 44 $ & BOSS CMASS & \protect\citet{Reid12}\\
	0.60 & $2234 \pm 115 $ & WiggleZ & \protect\citet{Bla11b}\\
       \hline
        0.35 & $1356 \pm 25$ & SDSS LRG & \protect\citet{Padmanabhan2012}\\
        0.57 & $2094 \pm 34$ & BOSS CMASS & \protect\citet{Aadvark:2012}\\
    \end{tabular}
  \end{center}
  \caption{Constraints on $D_V$  from recent surveys shown in
  Fig.~\ref{fig:dvfm}. We do not include the BAO measurements of
  \citet{Padmanabhan2012} or \citet{Aadvark:2012} in our analysis, as
  they are correlated with the measurements of \citet{Rei10,Reid12}.}
  \label{table:Dvconstraints}
\end{table}

\begin{table}
  \begin{center}
    \begin{tabular}{llll}
      {\it z} & $F$ & survey & reference\\
        \hline
	0.22 & $0.28 \pm 0.04$ & WiggleZ & \protect\citet{Bla11c}\\
	0.41 & $0.44 \pm 0.07$ & WiggleZ & \protect\citet{Bla11c}\\
	0.57 & $0.67 \pm 0.026$ & BOSS CMASS & \protect\citet{Reid12}\\
	0.60 & $0.68 \pm 0.06$ & WiggleZ& \protect\citet{Bla11c}\\
	0.78 & $0.97 \pm 0.12$ & WiggleZ& \protect\citet{Bla11c}\\
      \end{tabular}
  \end{center}
  \caption{AP constraints on $F$  from recent surveys shown in
  Fig.~\ref{fig:dvfm}.}
  \label{table:Hconstraints}
\end{table}

The bands on Fig.~\ref{fig:dvfm} show theoretical predictions of the
spatially-flat $\Lambda$CDM model when a WMAP 7 prior is applied to
the relevant cosmological parameters, and the Einstein-DeSitter (EdS)
model ($\Omega_{\rm m} = 1$) with $H_0 = 73.8 \pm 2.4 {\rm km/Mpc}/s$
(as measured by \citealt{riess:2011}).

\section{Modelling deviations from $\Lambda$CDMGR}
\label{sec:theory}

The standard cosmological model comprises three main assumptions: first, the
Universe on large scales is homogeneous and isotropic, secondly, DE is Cosmological
Constant and thirdly, gravity is described by GR on all scales. In this model the
background geometry of the Universe can be fully described by three numbers that
can be chosen to be the current relative energy densities of nonrelativistic
matter $\Omega_{\rm m}$ and cosmological constant $\Omega_\Lambda$ and the
current value of Hubble expansion rate $H_0$. The angular diameter distance and
Hubble expansion rate at any redshift are given by 
\begin{align}
  D_A(z) & = \frac{c}{H_0(1+z)}\chi\left(\displaystyle\int_0^z\frac{dz'}{E(z')}\right),\\
  H(z) & = H_0E(z),
  \label{eq:DAH}
\end{align}
\noindent
where
\begin{equation}
  \chi(x) = \left\{
  \begin{array}{l l}
    x &\mbox{if }\Omega_{\rm k}=0\\
    \frac{1}{\sqrt{\Omega_{\rm k}}}\sinh\left(\sqrt{\Omega_{\rm k}}x\right) &
    \mbox{if } \Omega_{\rm k}>0\\
    \frac{1}{\sqrt{-\Omega_{\rm k}}}\sin\left(\sqrt{-\Omega_{\rm k}}x\right) &
    \mbox{if } \Omega_{\rm k}<0
  \end{array} \right.,
\end{equation}
\noindent
the quantity $\Omega_{\rm k} = 1-\Omega_{\rm m}-\Omega_\Lambda$ is the relative energy density of spatial curvature and 
\begin{equation}
  E(z) = \sqrt{\Omega_m(1+z)^3 + \Omega_{\rm k}(1+z)^2 + \Omega_\Lambda}.\\
  \label{eq:Ez}
\end{equation}

When the three basic cosmological parameters are fixed the growth of matter
overdensities in GR to linear order in overdensities follows the growth equation
\begin{equation}
  \frac{d^2G}{d\ln a^2}+\left(2+\frac{d\ln H}{d\ln a}\right)\frac{dG}{d\ln a} =
  \frac{3}{2}\Omega_{\rm m}(a)G,
  \label{eq:growthrate}
\end{equation}
where $a=1/(1+z)$ is the scale factor and the growth factor
$G(a)=\delta(a)/\delta(a_{\rm ini})$ shows by how much the overdensities have
grown compared to some arbitrary initial time $a_{\rm ini}$. In a spatially-flat
$\Lambda$CDMGR model the growth rate $f = d \ln G/d \ln a$ can be well
approximated by a fitting formula $f(a) = \Omega_{\rm m}(a)^{0.55}$
\citep{Peebles1980}.

\subsection{Parametrizing deviations from GR}
\label{ssec:mgtheory}

A large number of viable alternatives to GR have been suggested within
the scientific community, each with its own theoretical justifications and advantages \citep[for
a recent review of MG see, e.g.,][]{Clifton:2012}. The deviations from GR are usually
constrained in terms of phenomenological parameterisations, with two main
approaches, either parametrizing the observables (such as $f$ and $G$)
or the perturbation equations (Eq.~\ref{eq:growthrate}).

The most widely used parameterisation for the growth rate $f$ allows
for deviations from GR through \citep{Wang:1998}
\begin{equation}
  f = \Omega_{\rm m}(a)^\gamma,
  \label{eq:fgamma}
\end{equation}
\noindent
where $\gamma$ is a free parameter that mildly depends on background geometry.
In GR this dependence can be approximated well by 
\begin{align}
  \gamma = 0.55 + 0.05[1 + w(z=1)]\ {\rm if } \ w>-1,\\
  \gamma = 0.55 + 0.02[1 + w(z=1)]\ {\rm if } \ w<-1.
  \label{eq:gamma}
\end{align}
\noindent
\citep{Linder:2005}. More accurate expressions for $\gamma$ as a function of
cosmological parameters and redshift that take into account higher order terms
in $1-\Omega_{\rm m}$ and curvature and are correct at sub-percent level have
been proposed \citep[see e.g.,][]{Linder2007,Ishak2009}. In GR we expect to find
a value of $\gamma\simeq0.55$. Higher values of $\gamma$ would be indicative of a
slower growth of structure than in GR and vice versa.

An alternative way of parameterising deviations from GR is to phenomenologically
modify the linear equations for dark matter perturbations instead of
parametrizing the solutions as in Eq.~(\ref{eq:fgamma}). To describe the most
general deviations from GR two scalar functions are necessary to modify the
temporal and spatial parts of perturbed Einstein's equations and many different
choices for these functions have been proposed in literature \citep[see
e.g.,][]{Bertschinger2006,Caldwell2007,Amendola2008,Amin2008}.

We consider the parameterisation in which the Eq.~(\ref{eq:growthrate}) is
modified to 
\begin{equation}
  \frac{d^2G}{d\ln a^2}+\left(2+\frac{d\ln H}{d\ln a}\right)\frac{dG}{d\ln a} =
  \frac{3}{2}\Omega_{\rm m}(a)G\mu(a),
  \label{eq:growthratemg}
\end{equation}
where $\mu(a)$ parametrizes the time-dependent deviation from GR and
could also be a function of scale. Eq.~(\ref{eq:growthratemg}) is valid when
the dark matter is cold and DE does not cluster on the scales of interest.  The
function $\mu(a)$ is expected to be equal to unity in GR at all times and, inspired by the DGP
model, the time dependence of $\mu(a)$ is sometimes parametrized by
\begin{equation}
  \mu(a) = 1 + \mu_{\rm s} a^s.
  \label{eq:mus}
\end{equation}
Large values of $\mu_{\rm s}$ correspond to a stronger
time dependence for the effective Gravitational Constant \citep[see
e.g.,][]{Song2011,zhao2011}. Parameter $s$ describes the time-evolution of the
modification. The $s=0$ case corresponds to a
simple rescaling of the Newton Constant with $\mu_0 = (G_{\rm eff} - G_{\rm
GR})/G_{\rm GR}$ while for $s>0$, the deviation from GR is time dependent and
becomes more pronounced at late times. Larger values of $s$ correspond to
increasingly recent deviations from GR; therefore $\mu_{\rm s}$ is less well
constrained for larger values of $s$.

Besides modifications to Eq.~(\ref{eq:growthrate}), MG theories can also
induce a gravitational slip $\eta=\Phi/\Psi$, which is a ratio of longitudinal
to Newtonian scalar potentials;\footnote{In Newtonian gauge $\Phi = \delta
g_0^0/2$ and $\Psi = \delta g_i^i/2$, where $\delta g_\mu^\nu$ is a perturbation
to the metric tensor.} in the absence of anisotropic sources, $\eta$ is
expected to be equal to unity \citep[see, e.g.,][]{Daniel2009}. The growth rate
data alone are not sensitive to gravitational slip and would have to be combined
with an observable such as weak lensing observations, that depend on $\Psi$,  to
yield constraints on both $\mu_s$ and $\eta$ functions simultaneously \citep[see
e.g.,][]{Song2011,zhao2011}.

If the amplitude of density perturbations is known accurately at some
earlier time (for example from CMB temperature anisotropy data at
recombination), and the background expansion history is relatively well known,
the growth rate measurements at low redshifts can be used to strongly constrain
deviations from GR. Even small modifications to the strength of gravitational
interaction can build up and result in significantly different predictions for
$f(z)\sigma_8(z)$ at late times; the constraints are stronger for models in
which modifications set in earlier.

\begin{figure}
  \includegraphics[width=90mm]{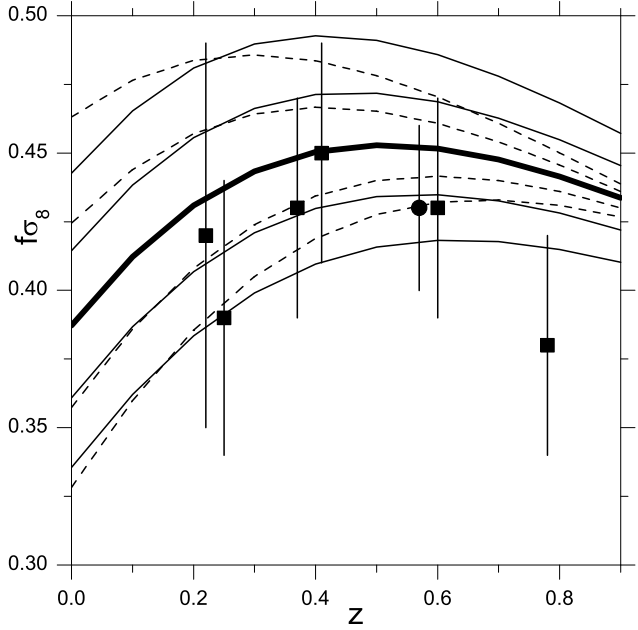}
  \caption{The data points show some of the measurements presented in
  Table~\ref{table:fs8constraints} with $1\sigma$ errorbars; the circle denotes
  the most recent CMASS DR9 measurement. The thick solid line shows the
  GR prediction, the thin solid lines show predictions for $\gamma = $ 0.45, 0.50, 0.60 and 0.65
  (top to bottom), and the thin dashed lines show predictions for $\mu_1 =$ 0.2, 0.1,
  -0.1 and -0.2 (top to bottom). $\Omega_{\rm m}(z=0)$ and $\sigma_8$ at recombination
  are kept fixed to their WMAP7 best-fit values.}
  \label{fig:effectmg}
\end{figure}

Fig.~\ref{fig:effectmg} shows $f(z)\sigma_8(z)$ predictions for
different MG models. To help to visualise the constraining power of
current RSD measurements some of the measurements presented in
Table~\ref{table:fs8constraints} are also plotted.

\subsection{Parametrizing deviations from $\Lambda$}
\label{ssec:wwtheory}

Deviations from a Cosmological Constant are most often parametrized by
$w_0$ and $w_{\rm a}$, where the time dependence of the EoS of DE is
approximated as
\begin{equation}
  \frac{p_{\rm DE}}{\rho_{\rm DE}} = w_0 + w_a\frac{z}{1+z}
  \label{eq:wz}
\end{equation}
\noindent
\citep{Chevallier2001,Linder2003} where $w_0 = -1$ and $w_{\rm a}= 0$ give the
$\Lambda$CDM limit. In this parameterisation, Eq.~(\ref{eq:Ez}) is modified to
\begin{equation}
  E(z) = \sqrt{\Omega_m(1+z)^3 + \Omega_{\rm k}(1+z)^2 + \Omega_\Lambda W(z)},
  \label{eq:wEz}
\end{equation}
where
\begin{equation}
  W(z) = (1+z)^{3(1 + w_0 + w_{\rm a})}\exp\left(\frac{-3w_{\rm a}z}{1 + z}\right).
  \label{eq:Wz}
\end{equation}

\begin{figure*}
  \includegraphics[width=180mm]{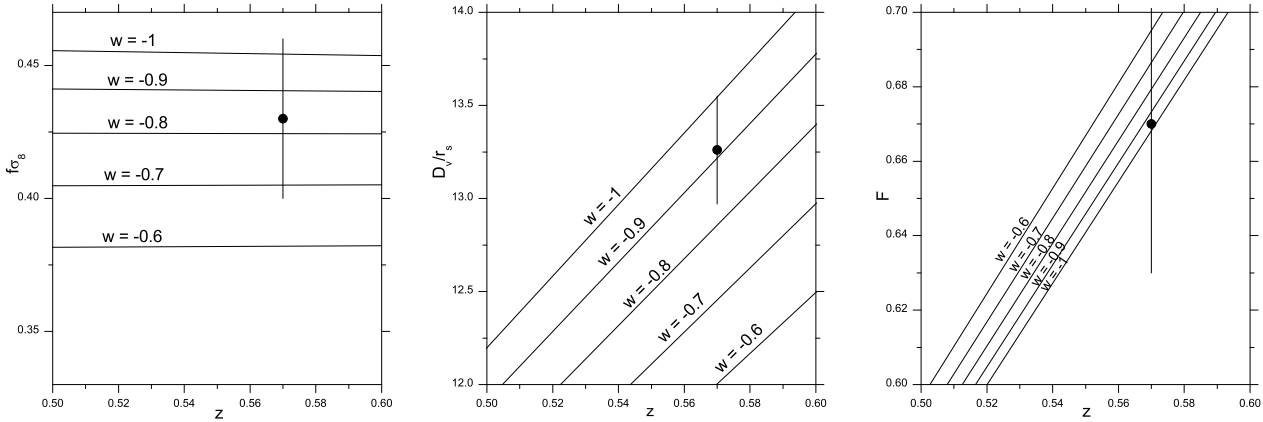}
  \caption{The data points show the estimates of $f(z)\sigma_8(z)$, $D_{\rm
  V}/r_{\rm s}$ and $F$ from \citet{Reid12} with $1\sigma$ errorbars. Solid
  lines correspond to theoretical predictions for different values of $w_0$ when
  $w_{\rm a}$ and $\Omega_{\rm k}$ are set to zero while the distance to the last
  scattering surface and $\sigma_8$ at recombination are kept fixed to their
  wMAP7 best-fit values.}
  \label{fig:effectw}
\end{figure*}

The change to the DE model modifies both the background expansion and the growth
of perturbations with respect to the $\Lambda$CDMGR model. This allows us to use
both geometric and growth rate measurements to constraint the model parameters.
To illustrate the sensitivity of \citet{Reid12} measurements of geometry and
growth to DE model parameters we plot them on fig.~\ref{fig:effectw} along with
theoretical predictions for different values of $w_0$.  To derive theoretical
predictions we set $\Omega_{\rm k}$ and $w_{\rm a}$ to zero and fixed the
distance to the CMB and amplitude of fluctuations at the last scattering
surface. In reality each line on fig.~\ref{fig:effectw} corresponds to a band
due to uncertainties in remaining cosmological parameters such as $\Omega_{\rm
m}h^2$. When these uncertainties are properly taken into account the absolute
error in $w_0$ resulting from $D_{\rm V}(z = 0.57)/r_{\rm s}$ is of order of
$0.24$ instead of $0.1$ as a naive inspection of fig.~\ref{fig:effectw} may
suggest. Fig.~\ref{fig:effectw} also shows that changing the value of $w_0$ has
an opposite effect on $f\sigma_8$ and $F$. Since the actual measurements are
strongly positively correlated their combination results in a much stronger
constraint on $w_0$ than what their individual marginalised errors imply.

\subsection{Constraining the DE potential}
\label{ssec:qtheory}

A more general way of describing deviations from $\Lambda$ is to constrain an
effective Lagrangian of the field responsible for the accelerated expansion
\citep[see e.g.,][]{Jimenez2011,Bloomfield2011}. If we assume that DE can be
effectively described by a scalar field that is minimally coupled to
gravitation, the Cosmological Constant will correspond to a scalar field with
flat potential $V(\phi) = {\rm const}$.  Observational data can be used to
constrain higher order terms in a Taylor expansion around the flat potential
\citep[see e.g.,][]{Jimenez2012}.

Instead of considering a general potential $V(\phi)$ as in \citet{Jimenez2012}
we look at specific scalar field dark energy models. We study two representative
theories: a slowly-rolling scalar field that has a tracker solution
\citep{Peebles:1988} and a pseudo Nambu-Goldstone boson (PNGB) in an oscillating
potential \citep{Frieman:1995}. Both of these models can result in a late-time
accelerated expansion of the Universe and have been shown to be compatible with
current data \citep[see e.g.,][]{samushiathesis,Dutta:2007}.\footnote{For a
recent review of scalar field DE models see e.g., \citet{Linder:2008}.} 

If the scalar field is minimally coupled to gravity its evolution is governed
by a Klein-Gordon type equation in the expanding Universe,
\begin{equation}
  \ddot{\phi} + 3H(z)\dot{\phi}-\frac{dV(\phi)}{d\phi} = 0,
  \label{eq:fddot}
\end{equation}
\noindent
where $V(\phi)$ is the self-interacting potential and the time-dependent energy
density of the scalar field,
\begin{equation}
  \rho_\phi = \frac{1}{2}\dot{\phi}^2 + V(\phi),
  \label{eq:omegaphi}
\end{equation}
\noindent
contributes to the Friedmann equations.

For the slowly-rolling tracker field, following \citet{ratra88}, we will
consider an inverse power-law self-interacting potential of the form
$V(\phi)\propto \phi^{-\alpha}$ and will assume that it does not cluster on the
scales of relevance. The quantity $\alpha$ is a positive constant that parametrizes the
steepness of scalar field potential. In the limit of $\alpha\rightarrow0$, the
potential becomes flat and the model approaches a cosmological constant; higher
values of $\alpha$ correspond to stronger time-evolution. The $\alpha$ parameter
has been constrained before using compilation of data sets and found to be less
than 0.7 at the $1\sigma$ confidence level \citep{samushiathesis}.

For the PNGB, following \citet{Frieman:1995}, we will consider an oscillating
potential of the form $V=M^4\left[1+\cos(\phi/f)\right]$, where $M$ and $f$ are
free parameters related to the energy scales of explicit and spontaneous symmetry
breaking respectively. This model possesses technical naturalness and can
naturally incorporate both the small value of cosmological constant and the
small mass of the scalar field today. Theoretical predictions of this model have
been compared to data \citep[see e.g.,][]{Waga:2000,Ng:2001,Kawasaki:2001}; to
be consistent with current observations the PNGB field has to be sitting on the
top of oscillatory potential or rolling down very slowly and the mass of the
scalar field must be larger than $\sim M_{\rm pl}/3$ \citep{Dutta:2007}.

Even though the dynamics of scalar field models can often be very well described
by a pair of $w_0$, $w_{\rm a}$ parameters, fitting data directly to physical
models has several advantages. The $w_0$-$w_{\rm a}$ parameterisation implicitly
imposes a theoretical prior that all values of those parameters are equally
likely. By directly examining specific DE models we can include the information
that some values of $w_0$ and $w_{\rm a}$ are easier to procure than the others.
This approach also sometimes reduces the number of independent parameters,
improving the constraining power of the data.

\subsection{Model-independent properties of DE}
\label{ssec:recon}

The $\Lambda$CDM paradigm leads to a number of interesting conclusions
about the late time Universe:

\begin{itemize}
  \item The expansion of the Universe must be accelerating.
  \item The Universe at present must be dominated by DE; the contribution of
    nonrelativistic matter being subdominant.
  \item The Cosmological Constant must be a relatively recent phenomenon,
    quickly declining in importance at higher redshifts.
\end{itemize}

Whether or not Universe exhibits these properties can be checked with
observational data without referring to a specific DE model.

The acceleration of the kinematic expansion of the Universe can be described by
the ``deceleration parameter'' $q(z)$, which is defined as
\begin{equation}
  q(z) \equiv - \frac{\ddot{a}a}{\dot{a}^2} = \frac{d\dot{a}(z)}{dz}\frac{1+z}{\dot{a}(z)}.
  \label{eq:qz}
\end{equation}
Negative values of $q$ indicate an accelerated expansion and vice versa. 

The ${\rm Om}$ parameter, defined as 
\begin{equation}
  {\rm Om}(z) \equiv \frac{\left[H(z)/H_0\right]^2-1}{(1+z)^3-1} =
  \frac{\dot{a}^2(1+z)^2/H_0^2-1}{(1+z)^3-1},
  \label{eq:om}
\end{equation}
\noindent
is expected to be constant (equal to the present day matter density $\Omega_{\rm m}$) in spatially-flat
$\Lambda$CDM Universe \citep{Sahni:2008} and can also be used to test
acceleration even if the value of $\Omega_{\rm m}$ is not known. In
a universe dominated by DE, $\Omega_{\rm m}$ is expected to be less than the
critical value irrespective of DE model.

Finally, the fact that DE becomes relevant very suddenly at low redshifts can be
tested by looking at the ratio of energy densities of DE to nonrelativistic
matter
\begin{equation}
  \frac{\rho_{\rm DE}(z)}{\rho_{\rm m}(z)} = \frac{H(z)^2/H_0^2 - \Omega_{\rm
  m}(1+z)^3}{\Omega_{\rm m}(1+z)^3}.
  \label{eq:fracDEz}
\end{equation}

In Sec.~\ref{ssec:reconstr} we will use the geometric part of our
measurements in combination with previous similar measurements tabulated in
Tables~\ref{table:Dvconstraints} and \ref{table:Hconstraints} to reconstruct these
model-independent quantities.

\section{Measuring deviations from $\Lambda$ and GR}
\label{sec:constraints}

We use the correlated estimates of $f\sigma_8$, $D_{\rm V}/r_{\rm s}$ and $F$
from \citet{Reid12} to constrain parameters describing deviations from
$\Lambda$CDMGR. We use the Monte Carlo Markov Chain \citep[MCMC; see
e.g.,][]{Tanner1996} method to explore likelihood surfaces. 
We do this by making use of publicly available {\sc cosmomc} code 
\citep{Lewis2002} that uses {\sc camb} to compute power-spectra
for the CMB and matter fluctuations \citep{Lewis2000}. The {\sc CosmoMC}
software was modified to support a time-dependent DE EoS \citep{Fang2008}, the
SNLS dataset \citep{Conley2011}, the measurements of growth and geometry from
\citet{Reid12} and general measurements of growth rate.

\subsection{Constraining deviations from GR}

For modified gravity we measure parameters $\gamma$, or $\mu_{\rm s}$
and $s$, as
described in Section~\ref{ssec:mgtheory}. We assume that the background evolution of
the Universe follows the predictions of spatially-flat $\Lambda$CDM with basic
cosmological parameters additionally constrained by SNeIa
\citep{Conley2011} and CMB \citep{Komatsu2011} data. We also combine
the measurements of \citet{Reid12} with previous measurements of the growth factor.
These measurements are somewhat heterogeneous. Most were made by assuming a
fixed fiducial cosmological model, while \citet{SamPerRac11} and \citet{Reid12}
marginalize over the shape. Models with varying levels of complexity have been
used to address the problem of nonlinear contamination from structure growth and
real to redshift space mapping. In this section we will ignore these differences
and will treat the reported measurements and errorbars as a Bayesian likelihood
for $f\sigma_8$ at the redshift of interest. 

\begin{figure}
  \includegraphics[width=90mm]{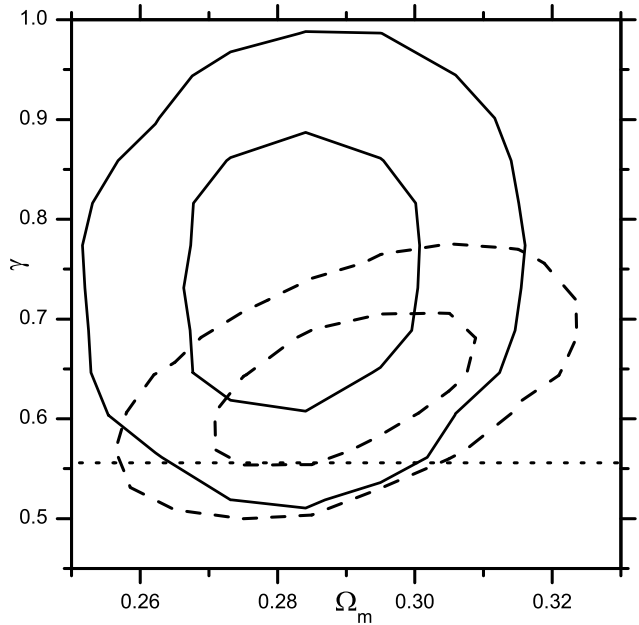}
  \caption{Confidence levels ($1\sigma$ and $2\sigma$) for joint fits to parameters $\gamma$
  and $\Omega_{\rm m}$. Solid lines show results for CMASS DR9 estimate of growth
  rate only, while dashed lines show results with previous estimates of growth
  rate added. In both cases the growth rate measurements are combined with CMB and SNeIa
  data. The dotted line shows the expected value in GR.}
  \label{fig:gammaconst}
\end{figure}

Fig.~\ref{fig:gammaconst} shows two-dimensional constraints on parameters
$\gamma$ and $\Omega_{\rm m}$. Using only the CMASS DR9 measurement of growth
rate we obtain $\gamma = 0.75 \pm 0.09$ when $\Omega_{\rm m}$ and $H_0$ are
marginalised over; when previous measurements of the growth rate from
Table~\ref{table:fs8constraints} are added we find $\gamma = 0.64 \pm 0.05$, an
8 percent measurement.  The addition of previous measurements of the growth rate
brings the best fit value closer to the GR expected value of $\gamma = 0.55$. In
both cases GR is about $2\sigma$ from the best fit.  \citet{Hudson2012} were
able to get a slightly stronger measurement of $\gamma$ by combining RSD data
with a low redshift peculiar velocity data from \citet{Turnbull2012} and
\citet{Davis2011}. \citet{Rapetti2012} find a constraint on $\gamma$ at a
similar level of precision, although their best-fit value is closer to the GR
prediction.

\begin{figure}
  \includegraphics[width=90mm]{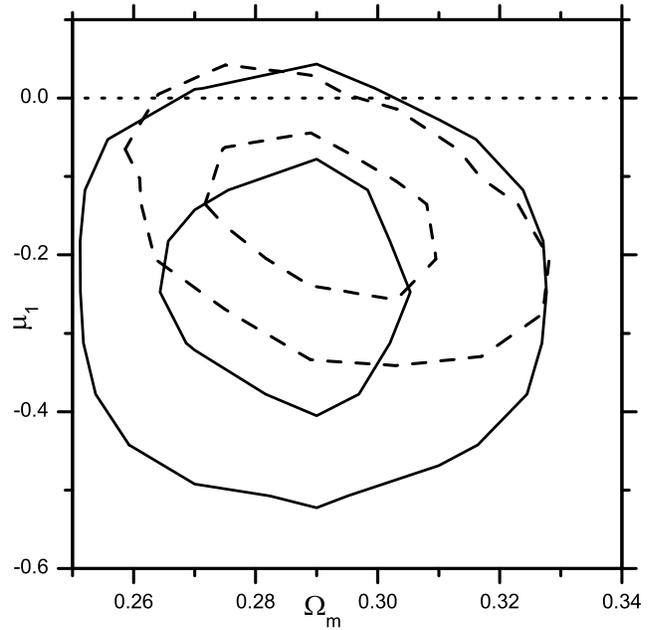}
  \caption{Confidence levels ($1\sigma$ and $2\sigma$) for joint fits to parameters  $\mu_1$ ($s
  = 1$) and $\Omega_m$. Solid lines show results for CMASS DR9 estimate of growth
  rate only, while dashed lines show results with previous estimates of growth
  rate added. In both cases the growth rate measurements are combined with CMB
  and SNeIa data. The dotted line shows the expected value in GR.}
  \label{fig:Ommu1}
\end{figure}

A similar two-dimensional likelihood contour plot of $\mu_1$ ($s=1$ is fixed)
and $\Omega_{\rm m}$ is presented in Fig.~\ref{fig:Ommu1}. With CMASS DR9
measurements we obtain $\mu_1 = -0.24 \pm 0.11$ when $\Omega_{\rm m}$ and $H_0$
are marginalised over; after adding previous measurements of growth rate the
constraint improves to $\mu_1 = -0.17 \pm 0.07$. The GR predicted value of
$\mu_1 = 0.0$ is again about $2\sigma$ from the best fit.

A summary of constraints on MG parameters is presented in
Table~\ref{table:mg}.

\begin{table}
  \begin{center}
    \begin{tabular}{lll}
      parameter & CMASS DR9 only & CMASS DR9 + other $f\sigma_8$\\
        \hline
	$\gamma$ & $0.75 \pm 0.09$ & $0.64 \pm 0.05$\\
	\hline
	$\mu_1$ & $ -0.24 \pm 0.11 $ & $-0.17 \pm 0.07$\\
      \end{tabular}
    \end{center}
    \caption{Constraints on parameters describing deviation from GR from CMB,
    SNeIa and growth rate data.}
    \label{table:mg}
\end{table}

The constraining power of the growth rate data is strongly augmented by the fact
that the initial amplitude of fluctuations at $z\sim 1000$ is very tightly
constrained by CMB data. This CMB prior on the initial amplitude of the density
fluctuations, which is degenerate with the optical depth to the last scattering
surface, allows us to convert our low redshift measurements of $f(z)\sigma_8(z)$
into pure measurements of growth rate $f(z)$ that has a much stronger
discriminative power. The detection of lensing effect on the power-spectrum of
temperature and polarisation anisotropies from ongoing {\it Planck} satellite
will enable a more robust determination of the amplitude of primordial
fluctuations \citep[see e.g.,][]{Zaldarriaga1998,Stompor1999,Amblard2004} which
in turn will further boost the constraining power of RSD measurements on MG
parameters.

One of the key assumptions in the \citet{Reid12} analysis is that physics at
recombination is very close to the ``standard'' model prediction, so that the
CMB prior on the shape and amplitude of the primordial fluctuations can be used.
This assumption breaks down for the $\mu_{\rm s}$ parameterisation, for low
values of $s$ that result in the rescaling of Newton's constant at earlier
times.  Modification of gravity at the last scattering surface will change the
likelihood of observed CMB anisotropies and will not be consistent with the
assumptions under which the measurements of \citet{Reid12} were derived. 

\begin{figure}
  \includegraphics[width=90mm]{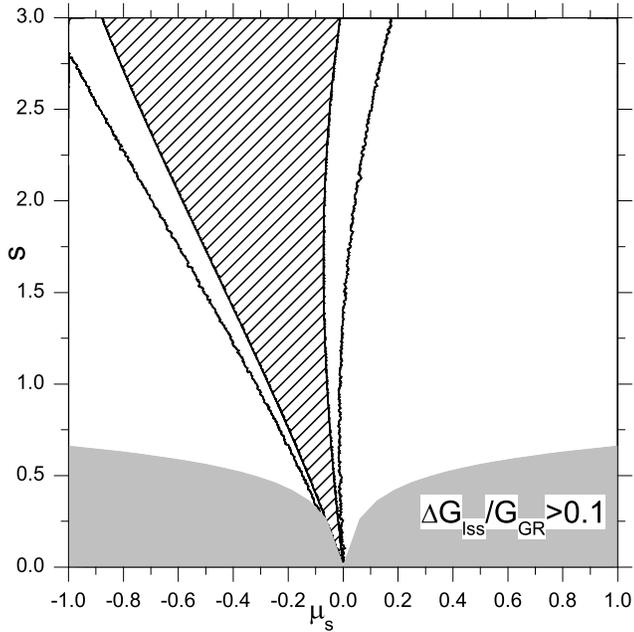}
  \caption{Confidence levels ($1\sigma$ and $2\sigma$) for joint fits to parameters $\mu_{\rm
  s}$ and $s$. Grey areas show parts of the parameter space for which the
  effective Newton's constant at last scattering surface is modified by
  greater than 10 per cent.}
  \label{fig:mus}
\end{figure}

Fig.~\ref{fig:mus} shows our constraints in $\mu$-$s$ space. Current
CMB data provides a 10 per cent constraint on Newton's constant at
high redshift, which is expected to be reduced to a 1.5 per cent
constraint with {\it Planck} data \citep{Galli2009}. The region of
parameter space excluded by this limit is shown by the grey shading in
Fig.~\ref{fig:mus}. For the remaining area, we ignore the effect of
the $<10$ per cent change in Newton's constant on the primordial power
spectrum.

Another important assumption in \citet{Reid12} is that the growth is scale
independent. In many MG theories such as $f(R)$, the growth of density
fluctuations is in fact strongly scale dependent \citep[see
e.g.,][]{Clifton:2012}. Interpreting the growth rate data in the light of these
models is not trivial since the measurements were made assuming a scale
independent growth (even though \citealt{Reid12} accounts for the scale dependence of
power-spectrum due to nonlinear evolution).  Testing these models with the
growth rate data would require refitting the model to the measured anisotropic
clustering while properly accounting for all the scale dependencies, which is
outside of the scope of this paper. Our results could be considered as
constraints at an effective scale, corresponding to those fitted by
\citet{Reid12}.

\subsection{Constraining deviations from $\Lambda$}

\begin{figure*}
  \includegraphics[width=180mm]{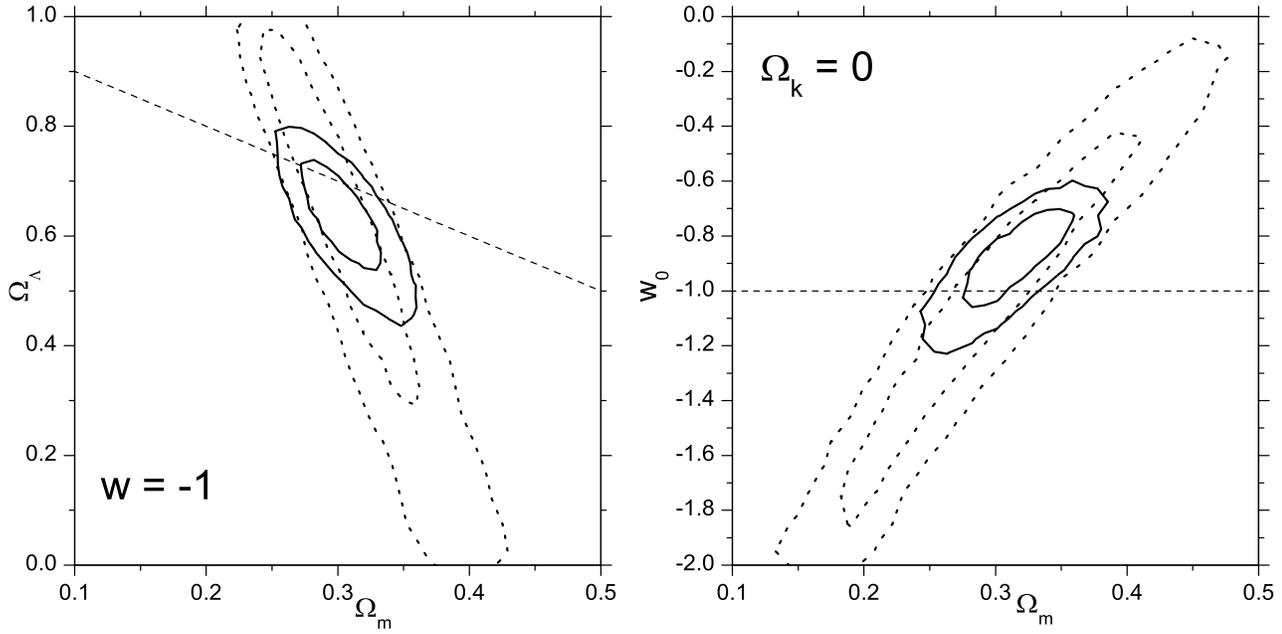}
  \caption{Confidence levels ($1\sigma$ and $2\sigma$) resulting from
    CMASS DR9 measurements and the WMAP7 shape and primordial
    amplitude prior. The solid lines show joint constraints from
    growth rate $f\sigma_8$ and geometrical measurements $D_{\rm V}$
    and $F$, while the dotted lines show constraints from geometrical
    measurements only. The dashed line on left panel denotes the
    spatially-flat Universe, while the same line on the right panel
    denotes a cosmological constant (i.e. both lines show the locus of
    $\Lambda$CDM models). In both panels, $w_{\rm a} = 0$.}
  \label{fig:omolw}
\end{figure*}

For DE we consider parameters $w_0$ and $w_{\rm a}$, as described in
Section~\ref{ssec:wwtheory}. We will define the following models: 
\begin{description}
  \item[$\Lambda$CDM,] where $w_0 = -1.0$, $w_{\rm a}=0.0$ and $\Omega_{\rm k} = 0.0$; 
  \item[wCDM,] where $w_0$ is free but $w_{\rm a}=0.0$ and $\Omega_{\rm k}=0.0$; 
  \item[$w_0w_{\rm a}$CDM,] in which both $w_0$ and $w_{\rm a}$ are free and $\Omega_{\rm k}=0.0$; 
  \item[O$\Lambda$CDM,] as $\Lambda$CDM but with free spatial curvature; 
  \item[OwCDM,] as wCDM but with free spatial curvature. 
\end{description}
To demonstrate the constraining power and degeneracy directions resulting from
our anisotropic measurements we first combine them with a WMAP7 prior on the
shape of the primordial power-spectrum and amplitude only. We do this by using
WMAP7 data to fit for a correlated joint likelihood of $\Omega_{\rm m}h^2$ and
the amplitude of matter fluctuations $\sigma_8$ at the CMB last scattering
surface, with other parameters marginalized over. As we keep the amplitude of
fluctuations fixed at the last scattering surface, the WMAP7 prior results in
different values of $\sigma_8(z=0)$ for different sets of parameters
$\Omega_{\rm m}$ and $w_0$. Fig.~\ref{fig:omolw}
shows resulting constraints for $\Lambda$CDM and wCDM models. 
For the O$\Lambda$CDM model we find $\Omega_{\rm m} = 0.308 \pm 0.022$ and $100\Omega_{\rm
k} = 5.9\pm 4.8$. In wCDM we obtain $w_0 = -0.91 \pm 0.12$.
Fig.~\ref{fig:omolw} clearly demonstrates how strongly complementary the
RSD measurement of growth rate are to the geometric measurements.  With our
geometric and growth measurements alone, supplemented by WMAP7 shape and
amplitude prior, we are able to constrain $\Omega_{\rm m}$ with a 7 per cent
precision and $w_0$ with a 14 per cent precision.

\begin{figure*}
  \includegraphics[width=180mm]{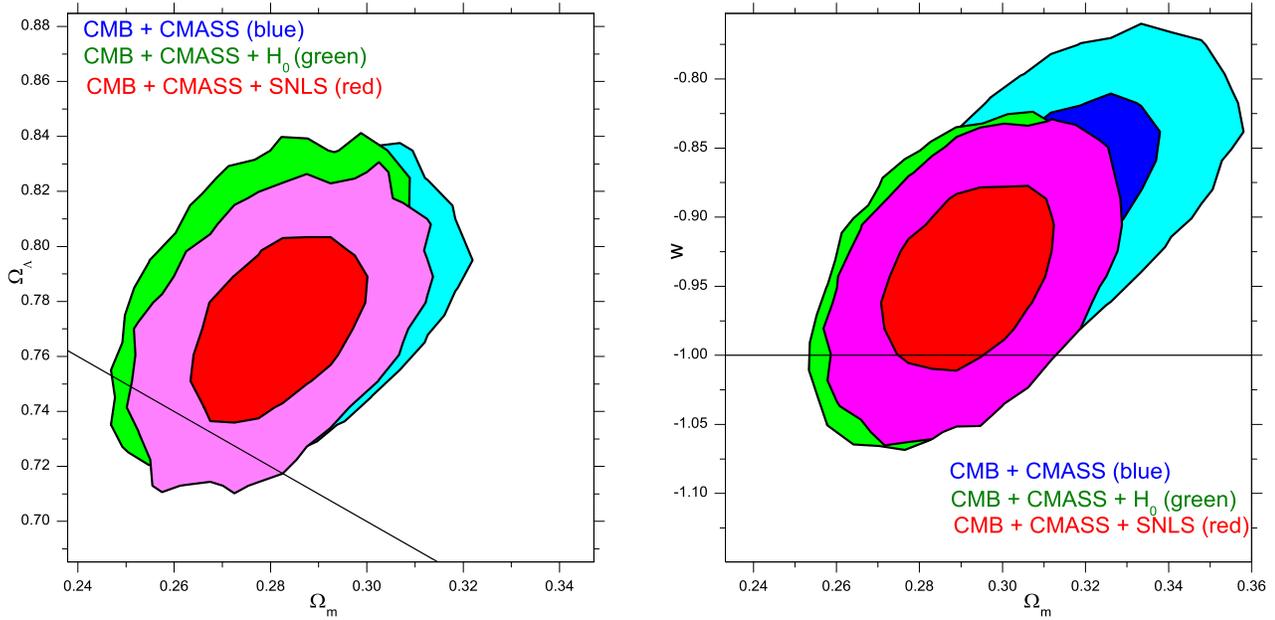}
  \caption{Confidence levels ($1\sigma$ and $2\sigma$) resulting from
    CMASS DR9 measurements combined with full WMAP7 likelihood, SNLS3
    data of SNeIa and the estimate of Hubble constant. The solid line
    on the left panel denotes the spatially-flat Universe, while the
    same line on the right panel denotes a cosmological constant
    (i.e. both lines show the locus of $\Lambda$CDM models). In both panels,
    $w_{\rm a} = 0$.}
  \label{fig:all}
\end{figure*}

Next, we combine BOSS measurements with the full WMAP7 likelihood, so that the 
angular diameter distance to the last scattering surface is also used.
To strengthen these constraints we will also combine the
CMASS DR9 data with SNeIa
\citep{Conley2011}, and $H_0$ \citet{riess:2011} data sets.
Table~\ref{table:all} presents the mean values and $1\sigma$ errors of
cosmological parameters with different model assumptions and combination of these data
sets. Fig.~\ref{fig:all} presents the results for O$\Lambda$CDM and wCDM models
when CMASS DR9 results are combined with full WMAP7 likelihood as well as either SNeIa 
or $H_0$ data.

\begin{figure}
  \includegraphics[width=90mm]{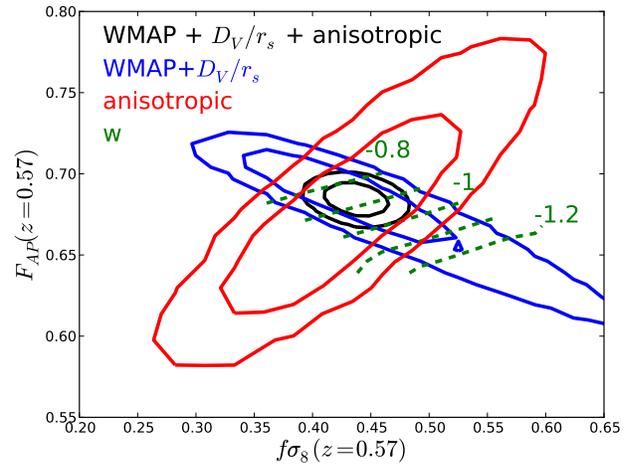}
  \caption{Constraints from \citet{Reid12} on the anisotropic galaxy clustering
parameters $f\sigma_8$ and $F_{AP}$ at $z=0.57$ are shown in red.  The blue
contour projects the combined constraints from WMAP and the spherically-averaged
BAO position from BOSS, $D_V/r_s$ at $z=0.57$; dashed green lines indicate the
mean value of $w$ in the region allowed by WMAP$+D_V/r_s$; the black contour
combines the anistropic measurement with the CMB and spherically-averaged BOSS
BAO position, resulting in a factor of $\sim 4$ improvement in the constraint on
$w$.}
  \label{fig:FAPfs8w}
\end{figure}

\begin{table*}
  \begin{center}
    \begin{tabular}{lllllll}
      \hline
      Cosmological model & Data set & $\Omega_{\rm m}$ & 1000$\Omega_{\rm k}$ & $H_0$ & $w_0$ & $w_a$ \\
        \hline
	$\Lambda$CDM& CMB + CMASS + SNeIa & 0.285 $\pm$ 0.014 & 0  & 68.9 $\pm$ 1.1 & -1 &  0\\
	\hline
	$\Lambda$CDM& CMB + CMASS & 0.291 $\pm$ 0.014 & 0 & 68.5 $\pm$ 1.2 &  -1 & 0\\
	\hline
	$\Lambda$CDM& CMB + CMASS + $H_0$ & 0.281 $\pm$ 0.013 & 0  & 69.5 $\pm$ 1.1 & -1 &  0\\
	\hline
	O$\Lambda$CDM& CMB + CMASS + SNeIa & 0.281 $\pm$ 0.014 & -9.2 $\pm$ 5.0  & 67.7 $\pm$ 1.3 & -1 &  0\\
	\hline
	O$\Lambda$CDM& CMB + CMASS & 0.288 $\pm$ 0.017 & -8.5 $\pm$ -5.4 & 67.4 $\pm$ 1.3 & -1 & 0\\
	\hline
	O$\Lambda$CDM& CMB + CMASS + $H_0$ & 0.277 $\pm$ 0.014 & -6.0 $\pm$ 4.9 & 68.8 $\pm$ 1.3 & -1 &  0\\
	\hline
	wCDM& CMB + CMASS + SNeIa & 0.292 $\pm$ 0.015 & 0  & 68.0 $\pm$ 1.4 & -0.94 $\pm$ 0.05&  0\\
	\hline
	wCDM& CMB + CMASS & 0.313 $\pm$ 0.017 & 0  & 65.9 $\pm$ 1.5 & -0.87 $\pm$ 0.05 &  0\\
	\hline
	wCDM& CMB + CMASS + $H_0$ & 0.291 $\pm$ 0.015  & 0  & 68.2 $\pm$ 1.4 & -0.93 $\pm$ 0.05 &  0\\
	\hline
	OwCDM& CMB + CMASS + SNeIa & 0.285 $\pm$ 0.017 & -8.2 $\pm$ 5.5 & 67.4 $\pm$ 1.5  & -0.98 $\pm$ 0.05 &  0\\
	\hline
	OwCDM& CMB + CMASS & 0.307 $\pm$ 0.022 & -3.9 $\pm$ 6.8 & 65.9 $\pm$ 1.6 & -0.90 $\pm$ 0.07 & 0\\
	\hline
	OwCDM& CMB + CMASS + $H_0$ & 0.285 $\pm$ 0.018 & -3.7 $\pm$ 5.7 & 68.2 $\pm$ 1.5 & -0.95 $\pm$ 0.07 & 0\\
	\hline
	$w_0w_{\rm a}$CDM& CMB + CMASS + SNeIa & 0.280 $\pm$ 0.018 & 0  & 68.8 $\pm$ 1.6 & -1.13 $\pm$ 0.12 & 0.65 $\pm$ 0.36 \\
	\hline
	$w_0w_{\rm a}$CDM& CMB + CMASS & 0.313 $\pm$ 0.037  & 0  & 66.2 $\pm$ 2.8 & -0.86 $\pm$ 0.34 &  -0.14 $\pm$ 1.04\\ 
	\hline
	$w_0w_{\rm a}$CDM& CMB + CMASS + $H_0$ & 0.261 $\pm$ 0.037 & 0  & 71.2 $\pm$ 2.3 & -1.29 $\pm$ 0.19 &  1.02 $\pm$ 0.48\\
	\hline
      \end{tabular}
    \end{center}
    \caption{Estimates of basic cosmological parameters for different models and
    combinations of data sets.}
    \label{table:all}
  \end{table*}

When we assume a cosmological constant and fix the spatial curvature to zero
($\Lambda$CDM), we obtain a 5 per cent constraint on $\Omega_{\rm m}$ and a
2 per cent constraint on Hubble constant from the joint fit to CMB data, CMASS
geometric measurements and SNeIa data. When the values of $w_{\rm a}$
and $\Omega_{\rm k}$ are fixed to zero, $w_0$ can be measured with a precision
of 5 per cent.

These results are consistent with the results reported in \citet{Aadvark:2012}
who use only the position of the BAO peak and \citet{Sanchez:2012} who use the
full shape of the monopole. Our measurements are derived from the same sample
but use additional information from the shape and the amplitude of the
quadrupole. This allows us to get significantly stronger constraints with the
CMASS data only (see Fig.~\ref{fig:omolw}). When CMASS data is combined with
other data sets this improvement is somewhat mitigated but still present. When
combined with CMB data and for the $\Lambda$CDM model our measurements improve
constraints on $\Omega_{\rm m}$ by about 18 per cent relative to the BAO only
results reported in \citet{Aadvark:2012}. The central values are consistent
within $1\sigma$. The biggest improvement is on the DE parameter $w_0$ ($-0.87
\pm -0.05$ compared with $-0.87 \pm
0.24$ in wCDM model), where we find a factor of four improvement over BAO only results even
after combining the BAO measurements with WMAP7. Fig.~\ref{fig:FAPfs8w}
illustrates why including the information from anisotropic BOSS clustering
improves the constraint on $w$ so much, even though Fig.~\ref{fig:effectw} shows
that the marginalized one-dimensional constraints on $f\sigma_8$ and the
Alcock-Paczynski parameter $F$ are individually not very constraining on $w$.
The anisotropic BOSS constraint (red) is nearly perfectly perpendicular to the
degeneracy direction opened by $w$ in the $f\sigma_8(z=0.57)-F_{AP}(z=0.57)$
plane for the data combination of WMAP7 with the BOSS measurement of
$D_V(z=0.57)/r_s$.  When combined with CMB data our measurements of growth and
geometry prefer $w_0>-1$ at about $2\sigma$ confidence level. When SNeIa or
$H_0$ data is added the agreement with
$\Lambda$CDM becomes better.

\subsection{Constraining the effective potential of scalar-field dark energy}
\label{ssec:qconstraints}

We also constrain parameters of scalar field dark energy models, as
described in Section~\ref{ssec:qtheory}, for spatially-flat
cosmological models.

\begin{figure*}
  \includegraphics[width=180mm]{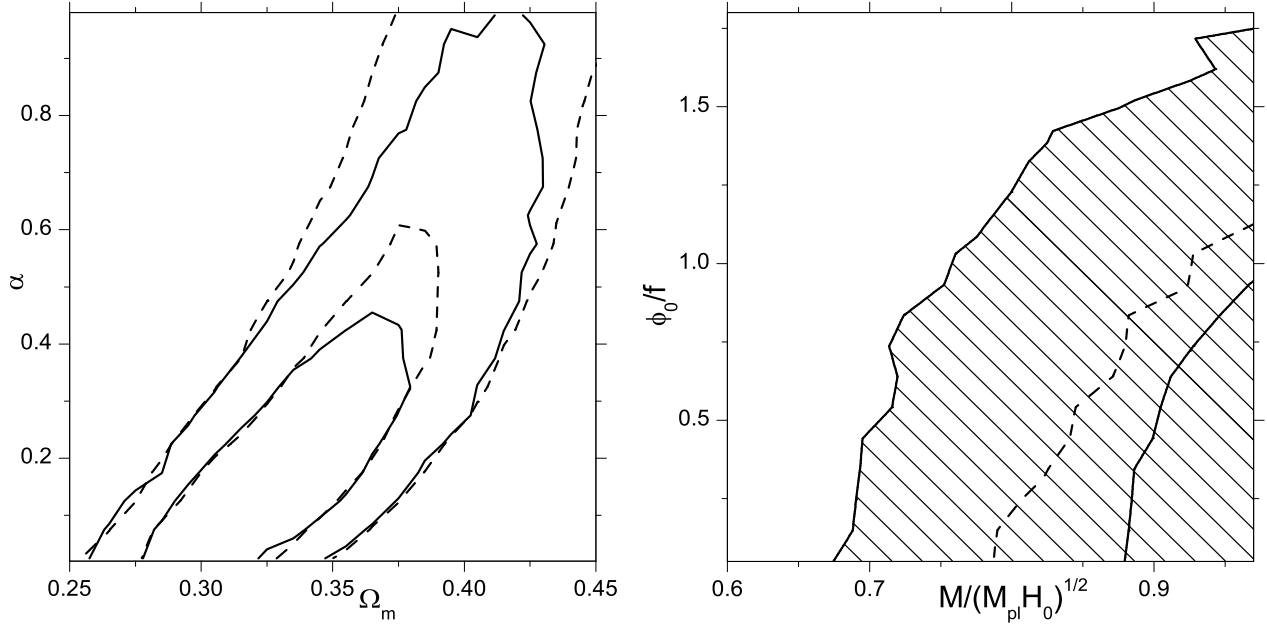}
  \caption{Confidence levels ($1\sigma$ and $2\sigma$) for joint fits to
  parameters $\Omega_m$ and $\alpha$ for the inverse-power law model (left
  panel) and parameters $M$ and $f$ for the PNGB model (right panel) adopting
  the CMASS DR9 measurements of $D_{\rm V}$ and $F$ and the WMAP7 shape prior.
  On the right panel, the shaded area corresponds to the region allowed at
  $2\sigma$. Without adding the growth rate information the PNGB constraints are
  weak. The dotted line shows $1\sigma$ contour from geometric measurements
  only, almost all of the remaining phase space is allowed at $2\sigma$.}
  \label{fig:scalar}
\end{figure*}

Fig.~\ref{fig:scalar} shows constraints on the parameter $\alpha$ describing the
potential of the slowly rolling scalar field and parameters $M$ and $f$
describing the potential of PNGB field from our geometric and growth
measurements combined with WMAP 7 shape and amplitude prior. For both models we
assume a spatially-flat Universe. We also assume a uniform theoretical prior
$0<f/\sqrt{M_{\rm pl}}<1$ and $M/\sqrt{M_{\rm pl}H_0}$, where $M_{\rm pl}$ is
the Planck mass.

The resulting constraints on basic scalar field parameters are presented in
Table~\ref{table:alpha}. In both cases the $\Lambda$CDM limit of the scalar
field theories provides a good fit to the data. The $\alpha$ parameter is
constrained to be less than 0.48 at $1\sigma$ confidence level, which is
slightly better compared to the previously reported best constraint
\citep{samushiathesis}. $M$ is needs to be at least 88 per cent of the Planck
mass. For comparison previous studies used SNeIa data to constrain $M$ to be
more than third of the Plank mass \citep{Dutta:2007}. The $\Lambda$CDM limits of
PNGB model ($\phi_0/f = 0.0$) and power-law model ($\alpha = 0$) provide a good fit to data.

\begin{table}
  \begin{center}
    \begin{tabular}{lcc}
      parameter & most likely value & $1\sigma$ confidence level \\
        \hline
	$\alpha$ & $0.16$ & $< 0.48$\\
	\hline
	$M/\sqrt{M_{\rm pl}H_0}$ & $ 1.0$ & $ > 0.9$\\
	\hline
	$\phi_0/f$ & $0.11$ & $< 1.17$\\
      \end{tabular}
    \end{center}
    \caption{Most likely values and $1\sigma$ confidence level constraints on
    parameters describing scalar field dark energy from our geometric and growth
    measurements and CMB shape and distance prior.}
    \label{table:alpha}
\end{table}

\subsection{Constraining model-independent properties of DE}
\label{ssec:reconstr}

To reconstruct the model-independent properties of the late-time Universe, as
described in Section~\ref{ssec:recon} we will use the geometric
\citet{Reid12} measurements in combination with the previous AP constraints presented in
Table~\ref{table:Hconstraints}. These estimates were derived using different
fitting methods, measurements and range of scales. We ignore these differences
in our analysis and will treat the reported measured values and their errorbars
as Bayesian likelihoods of $D_{\rm V}$ and $F$ at the redshifts of interest.

If we assume that the Universe is well described by a FRW metric, the
Hubble rate and expansion rate and the relationship with distances can be
written without a reference to a particular DE model; they can be expressed
purely as functions of $\dot{a(z)}$ through
\begin{align}
  D_A(z) = \frac{c}{(1+z)}\displaystyle\int_0^z\frac{dz'}{\dot{a}(z')(1+z')},
  \label{eq:Dadot}
  \\
  H(z)=\dot{a}(z)(1+z).  
  \label{eq:DHadot}
\end{align}
\noindent
where the scale factor $a$ is normalised such that $a=1$ today and dot denotes a time
derivative.

Different methods for the model-independent reconstruction of the expansion
history and properties of DE have been proposed in the past \citep[see
e.g.,][and references
therein]{Alam:2007,Turner:2007,Daly:2007,Shafieloo:2010,Crittenden:2012}. The
most popular approach is to parametrize the dimensionless coordinate distance in
the integral in Eq.~(\ref{eq:Dadot}) by a polynomial and express the scale factor
through its first and second derivatives.  \citet{Bla11c} used the method of
\citet{Shafieloo:2006} to reconstruct the comoving distance from SNeIa data,
which, combined with their measurements of $F$, enabled an estimate of $H(z)$. 

We will adopt an alternate approach that does not require taking derivatives of
the quantities reconstructed from data, which has been shown to introduce
artificial oscillations in reconstructed variables \citep{lazkoz2012}. We assume
that $a^3H^2(a)= a\dot{a}^2 \equiv y(a)$ is a smooth function of the scale factor, so that it
can be approximated by a third order polynomial. We have checked that this
approximation holds to a percent level for all conventional DE models for wide
range of redshifts; for $\Lambda$CDM this expansion is exact up to the radiation
dominated era. Eqs.~(\ref{eq:DHadot}) can be rewritten in terms of $y(a)$ as
\begin{align}
  D_A(a) = ca \displaystyle\int_a^1 \frac{a'^{3/2}da'}{\sqrt{y(a')}},
  \label{eq:Dy}
  \\
  H(a) = \sqrt{\frac{y(a)}{a^3}}.
  \label{eq:DHy}
\end{align}

We map $y(a) = \displaystyle\sum_{i = 0}^{3} y_i a^i$ into $D_{\rm V}$ and $F$
using Eqs~(\ref{eq:Dy}) -- (\ref{eq:DHy}) and compare the results to geometric
measurements. We compute a likelihood of polynomial coefficients of the
expansion of $\mathcal{L}(a_i)$ and map it onto the likelihood of deceleration
parameter $q(z)$ which is defined in Eq.~(\ref{eq:qz}) by
\begin{equation}
  \mathcal{L}^{\rm q}(q) = \mathcal{L}(y[a_i]).
  \label{eq:Lmap}
\end{equation}

\begin{figure*}
  \includegraphics[width=180mm]{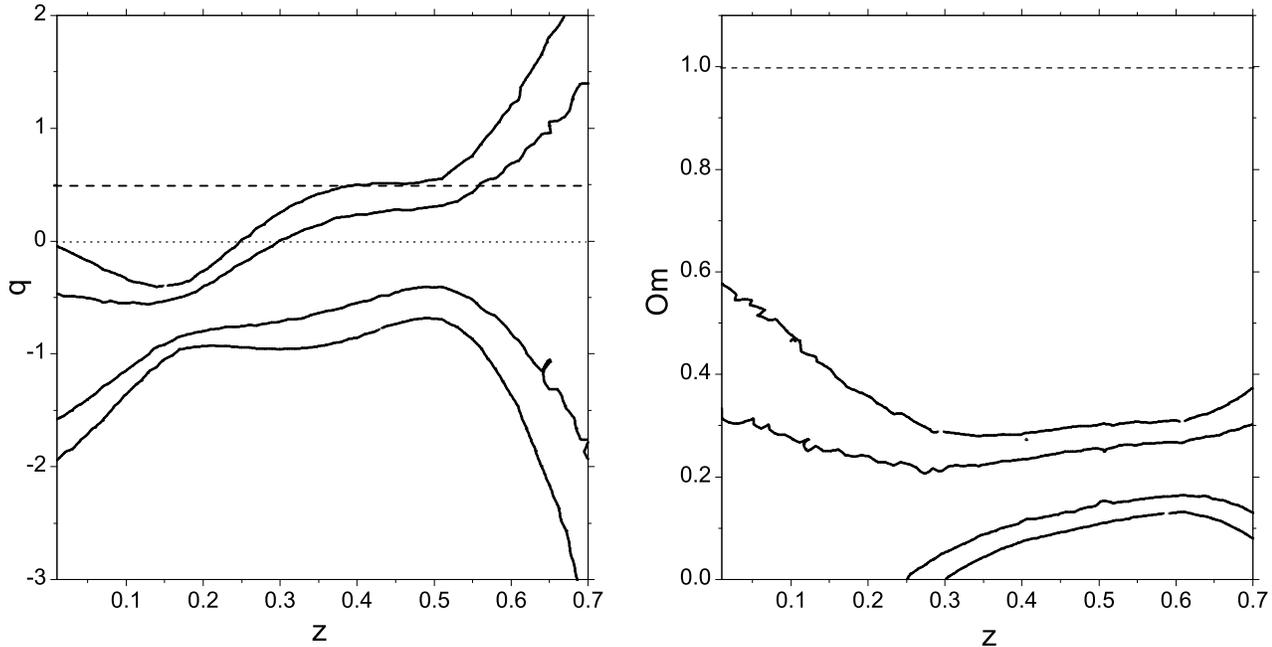}
  \caption{Confidence levels ($1\sigma$ and $2\sigma$) for the
    deceleration parameter as a function of redshift and ${\rm Om}(z)$
    reconstructed from the compilation of geometric measurements in
    tables~\ref{table:Dvconstraints} and
    \ref{table:Hconstraints}. $H_0$ is marginalised over with an HST
    prior. The dotted line in the left panel demarcates accelerating
    expansion (below the line) from decelerated expansion (above the
    line). The dashed line in both panels shows the expectation for an EdS model.}
  \label{fig:qom}
\end{figure*}

The left panel of Fig.~\ref{fig:qom} presents the results of the reconstruction
of $q(z)$ using our geometric measurements from the CMASS DR9 data combined with
previous similar geometric measurements from 6dFGRS, SDSS-II and WiggleZ
surveys. The reconstruction uses only the geometric measurements, shape prior
from WMAP7 data and theoretical priors of spatially-flat FRW metric and
smoothness of $a^3H^2(a)$. The recovered deceleration parameter is less than
zero at redshifts $z<0.3$ and more than $3\sigma$ from $q=0.5$ as predicted by
EdS model. \citet{Lampeitl2010} were able to derive a much stronger constraint
of $q = -0.34 \pm 0.18$ using SNeIa data only, however they had to assume
constant $q$ while we allow $q$ to have non-zero higher derivatives.

The right panel of Fig.~\ref{fig:qom} presents a similar reconstruction of
${\rm Om}(z)$ parameter.  The reconstructed  ${\rm Om}(z)$ is consistent with
being constant and is more than $5\sigma$ from the EdS predicted value of ${\rm
Om}(z) \equiv 1$. Our reconstructed $q(z)$ and ${\rm Om}(z)$ are consistent with
those reconstructed by \citet{Bla11c}; our results are smoother as a function
of redshift because of the different reconstruction method used.

Our geometric measurements can be used to derive an estimate of Hubble expansion rate
at $z = 0.57$, since
\begin{equation}
  H(z) = c\frac{(zF^2)^{1/3}}{D_{\rm V}}.
\end{equation}
\noindent
This measurement of $H(z)$ combined with an independent measurement of $H_0$ can
be used to estimate the increase in the fractional energy density of DE with
respect to nonrelativistic matter from $z = 0.57$ to present day. If we assume a
flat FRW background and two dominant components $\Omega_{\rm m}$ and
$\Omega_{\rm DE}$ then
\begin{equation}
  \frac{\rho_{\rm DE}(z)}{\rho_{\rm m}(z)} = \frac{H(z)^2/H_0^2 - \Omega_{\rm
  m}(1+z)^3}{\Omega_{\rm m}(1+z)^3}.
  \label{eq:rhoDEz}
\end{equation}
Combining our estimate of $H(z = 0.57)$ with the \citet{riess:2011} measurement of
$H_0$ and a WMAP 7 prior of $\Omega_{\rm m}h^2 = 0.1334 \pm 0.0056$ we find
$\rho_{\rm DE}(z = 0)/{\rho_{\rm m}(z = 0)} = 3.09 \pm 0.32$ and $\rho_{\rm
DE}(z = 0.57)/{\rho_{\rm m}(z = 0.57)} = 0.67 \pm 0.13$. This means that the
fractional contribution of DE to the total has increased by a factor of $\sim
4.7$ since $z = 0.57$, lending support to the notion that expansion dominated by
DE is a relatively recent occurrence.

\subsection{Is there an evidence for deviations from flat LCDMGR?} As discussed
in \citet{Reid12} the observed galaxy clustering when combined with CMB data is
fully consistent with LCDMGR expectations. Nevertheless, several of our fits
to one-parameter extensions of flat LCDMGR ($\gamma$, $\mu_1$, and $w_0$) indicate
a $\sim 2\sigma$ preference for values away from the fiducial. The data,
however, demand these extra parameters at a much lower significance than
$2\sigma$: the difference in the best fit $\chi^2$ values between flat LCDMGR
and these one-parameter model extensions is only $\sim$2.5. A similar effect has
been observed for the effective number of Neutrinos ($N_{\rm eff} = 3$ in the
standard model of particle physics), where the
Bayesian posterior likelihood resulting from the CMB data has been shown to
prefer $N_{\rm eff}>3$ at high confidence level. The \citet{Gon11}, however,
showed that in this case the priors imposed on cosmological parameters have a
strong effect on the posterior likelihood of $N_{\rm eff}$ and after removing
this prior-dependency the preference for the deviation from $N_{\rm eff}=3$ is
much lower.

\section{Conclusions}
\label{sec:conclusion}

We have used the \citet{Reid12} measurements of angular distance, Hubble
expansion rate and growth rate derived from the anisotropic clustering
of BOSS CMASS DR9 galaxies to place constraints on deviations from the
standard cosmological model that assumes a $\Lambda$CDM background
with structure formation driven by GR. The geometric measurements of
$D_{\rm V}$ and $F$ are complementary to similar measurements from the
BAO peak position \citep{Aadvark:2012} and the full shape of the
correlation function \citep{Sanchez:2012} and strengthen existing
constraints on parameters describing the time-dependence of DE energy
density. The RSD measurement of $f\sigma_8$ was shown to provide an
additional constraint on the parameters describing deviations from GR
and helped to significantly tighten DE constraints derived from
geometric measurements. We now highlight our findings by
using them to answer fundamental questions about our Universe.\\

\noindent{\it How much do RSD measurements enhance the geometric measurements?}\\
When GR is assumed the RSD measurements of growth break parameter degeneracies
present when using purely geometric measurements, and consequently significantly
tighten constraints on basic cosmological parameters. The addition of growth
rate information improves constraints on $\Omega_{\rm m}$ by 18 per cent
relative to the case where only BAO peak position data is used.  The improvement
is most dramatic for the $w_0$ parameter constraints, which improve by a factor
of four compared to BAO only results (see Table~\ref{table:all}).\\

\noindent{\it Does GR provide a good description of data?}\\
When $\Lambda$CDM is assumed our measurements of growth and geometry
show a $2\sigma$ preference for models in which the growth of
structure is weaker compared to GR. Adding previous RSD measurements
at other redshifts brings the best-fit closer to GR value but still
prefers somewhat weaker growth. This results from the fact that most
RSD measurements of $f(z)\sigma_8(z)$ with high signal-to-noise are below GR predictions 
(see Fig.~\ref{fig:fs8}).\\

\noindent{\it Does the Cosmological Constant provide a good description of data?}\\
Assuming GR, our measurements of growth and geometry show a $2\sigma$
preference for $w>-1$. When combined with SNeIa data, the best-fit is
closer to the cosmological constant model and the discrepancy is
lowered to about $1\sigma$. It should be noted that, for DE as well as
GR, the $2\sigma$ preference means a preference in terms of relative
Bayesian likelihood.\\

\noindent{\it How well can the DE scalar field potential be constrained?}\\
We demonstrated that our measurements of growth and geometry, when combined with
CMB information, provide strong constraints on scalar field DE model
parameters. The constraints obtained are better than previously reported
from different combinations of data sets. The flat potential
(Cosmological Constant) provides a good fit to data.\\

\noindent{\it Has the expansion of the Universe accelerated recently?}\\
We employed a model-independent approach that relies on very few general
assumptions to reconstruct the ``deceleration parameter'' at low redshifts. We
showed that current AP measurements provide $2$ to $3\sigma$ evidence for
the accelerated Universe at low redshifts. Much stronger measurements of $q(z)$
exist in the literature, but they depend on additional assumptions about the
nature of DE.\\

\noindent{\it Did DE emerge as a dominant component only very recently?}\\
Our estimate of $H(z=0.57)$ combined with an estimate of $H_0$ suggest that
energy density of DE relative to nonrelativistic matter was about 4.5 times
lower at $z=0.57$.\\

\noindent{\it Is the standard $\Lambda$CDMGR model still valid?}\\
Measurements of growth and geometry from CMASS DR9 sample allow for a
very strong test of MG and DE. When combined with WMAP7 data they show
a $2\sigma$ preference for either weaker gravity or $w>-1$. However,
in terms of absolute $\chi^2/{\rm dof}$, the simple $\Lambda$CDMGR
model still provides a good fit to the data.

\section*{Acknowledgements}

LS \& WJP are grateful for support by the European Research Council. LS
acknowledges partial support from SNSF SCOPES grant 128040 and GNSF grant
ST08/4-442. WJP also acknowledges support from the UK Science and Technology
Facilities Research Council. BAR gratefully acknowledges support provided by
NASA through Hubble Fellowship grant 51280 awarded by the Space Telescope
Science Institute, which is operated by the Association of Universities for
Research in Astronomy, Inc., for NASA, under contract NAS 5-26555.

Funding for SDSS-III has been provided by the Alfred P. Sloan
Foundation, the Participating Institutions, the National Science
Foundation, and the U.S. Department of Energy Office of Science. The
SDSS-III web site is http://www.sdss3.org/. 

SDSS-III is managed by the Astrophysical Research Consortium for the
Participating Institutions of the SDSS-III Collaboration including the
University of Arizona, the Brazilian Participation Group, Brookhaven
National Laboratory, University of Cambridge, Carnegie Mellon
University, University of Florida, the French Participation Group, the
German Participation Group, Harvard University, the Instituto de
Astrofisica de Canarias, the Michigan State/Notre Dame/JINA
Participation Group, Johns Hopkins University, Lawrence Berkeley
National Laboratory, Max Planck Institute for Astrophysics, Max Planck
Institute for Extraterrestrial Physics, New Mexico State University,
New York University, Ohio State University, Pennsylvania State
University, University of Portsmouth, Princeton University, the
Spanish Participation Group, University of Tokyo, University of Utah,
Vanderbilt University, University of Virginia, University of
Washington, and Yale University.

We acknowledge the use of the Legacy Archive for Microwave Background Data
Analysis (LAMBDA). Support for LAMBDA is provided by the NASA Office of Space
Science.  

Numerical computations were done on the Sciama High Performance Compute (HPC)
cluster which is supported by the ICG, SEPNet and the University of Portsmouth.

\appendix
\section{NonGaussianity in posterior likelihood}
\label{appendixA}

\begin{figure*}
  \includegraphics[width=180mm]{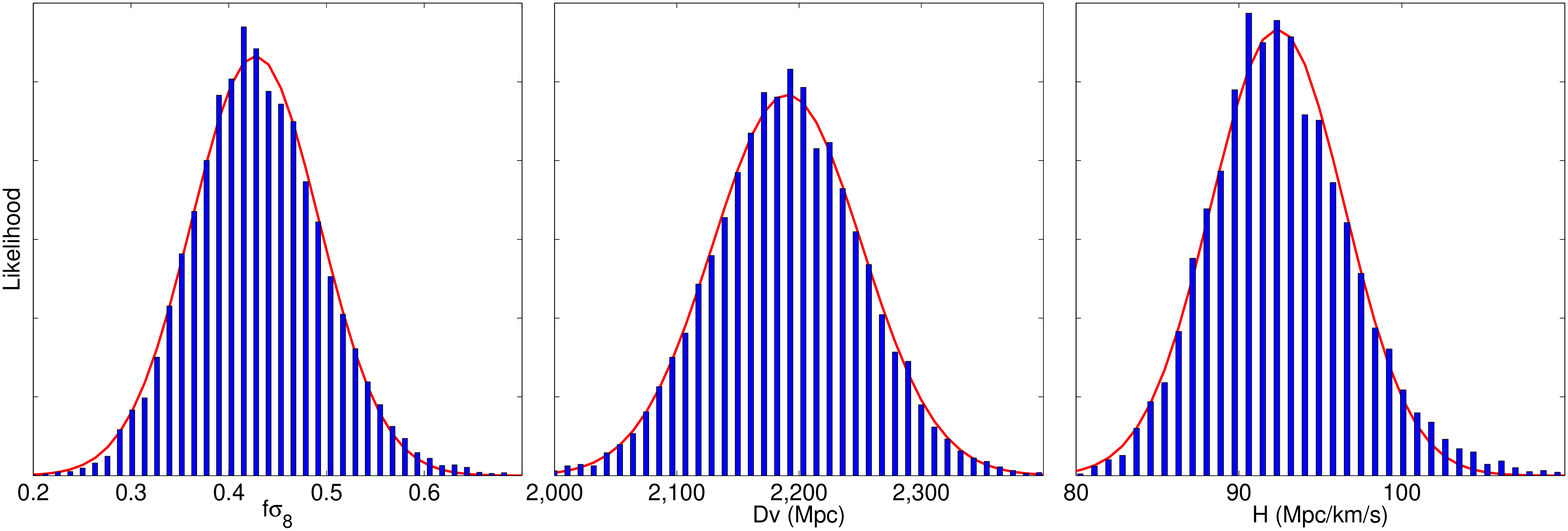} 
  \caption{Marginalized one-dimensional posterior likelihood of parameter
  $f\sigma_8$, $D_A$ and $H$ from \citet{Reid12} used in our analysis (The
  likelihood is in arbitrary units). The bars show the real distribution, while
  the solid line represents the Gaussian approximation that we use.}
  \label{fig:skewness}
\end{figure*}

Figure \ref{fig:skewness} shows the marginalized one-dimensional posterior
likelihood of parameters $f\sigma_8$, $D_A$ and $H$ used in our analysis. Our
multivariate Gaussian approximation to the likelihood works very well everywhere
except for the tails of the likelihood where it deviates from the empirical
likelihood by not accounting from for small positive skewness in $f\sigma_8$ and
$H$. We have checked that the derived likelihood of modified gravity and dark
energy parameters does not change at 1 and 2$\sigma$ confidence levels when the
Gaussian approximation is used instead of the whole likelihood.

\label{lastpage}


\begin{thebibliography}{999}

\bibitem[{{Amblard, Vale \& White}(2004)}]{Amblard2004}
  Amblard A., Vale C., White M., 2004, New Astron., 9, 687

\bibitem[{{Amendola, Kunz \& Sapone}(2008)}]{Amendola2008}
  Amendola L., Kunz M., Sapone D., 2008, Cosmology Astropart. Phys., 4, 13

\bibitem[{{Amin, Wagoner \& Blandford}(2008)}]{Amin2008}
  Amin M.A., Wagoner R.V., Blandford R.D., 2008, MNRAS, 390, 131

\bibitem[{{Anderson et al.}(2012)}]{Aadvark:2012}
  Anderson L., et al., 2012, preprint, [arxiv:1203.6594]

\bibitem[{{Alam, Sahni \& Starobinsky}(2007)}]{Alam:2007}
  Alam V., Sahni A., Starobinsky A., 2007, JCAP, 11, 0702

\bibitem[{{Albrecht et al.}(2009)}]{Albrecht2009}
  Albrecht A., et al., 2009, ``Findings of the Joint Dark Energy Mission Figure of
  Merit Science Working Group'', [arxiv:0901.0721]

\bibitem[{{Alcock \& Paczynski}(1979)}]{AP}
  Alcock C., Paczynski B., 1979, Nature, 281, 358.

\bibitem[{{da Angela et al.}(2008)}]{Ang08}
  da Angela J., et al., 2008, MNRAS, 383, 565

\bibitem[{{Appleby \& Linder}(2012)}]{Appleby2012}
  Appleby S.A., Linder E.V., 2012, JCAP, 08, 026

\bibitem[{{Bertschinger}(2006)}]{Bertschinger2006}
  Bertschinger E., 2006, ApJ, 648, 797

\bibitem[{{Beutler et al.}(2011)}]{Beutler11}
  Beutler F., et al., 2011, MNRAS, 416, 3017

\bibitem[{{Beutler et al.}(2012)}]{Beutler12}
  Beutler F., et al., 2012, MNRAS, 423, 3430

\bibitem[{{Blake et al.}(2011a)}]{Bla11a}
  Blake, et al., 2011a, MNRAS, 415, 2876

\bibitem[{{Blake et al.}(2011b)}]{Bla11b}
  Blake, et al., 2011b, MNRAS, 418, 1707

\bibitem[{{Blake et al.}(2011c)}]{Bla11c}
  Blake C., et al., 2011c, MNRAS, 418, 1725

\bibitem[{{Blake et al.}(2012)}]{Blake2012}
  Blake C., et al., 2012, MNRAS, 425, 405

\bibitem[{{Bloomfield \& Flanagan}(2012)}]{Bloomfield2011}
  Bloomfield J.K., Flanagan E.E., 2012, JCAP, 10, 039

\bibitem[{{Buchdahl}(1970)}]{Buchdahl1970}
  Buchdahl H.A., 1970, MNRAS, 150, 1

\bibitem[{{Cabre \& Gaztanaga}(2009)}]{Cabre:2009}
  Cabre A., Gaztanaga E., 2009, MNRAS, 393, 1183

\bibitem[{{Caldwell, Cooray \& Melchiorri}(2007)}]{Caldwell2007}
  Caldwell R., Cooray A., Melchiorri A., 2007, PRD, 76, 023507

\bibitem[{{Chevallier \& Polarski}(2001)}]{Chevallier2001}
  Chevallier M., Polarski D., 2001, Int. J. Mod. Phys. D, 10, 213

\bibitem[{{Chuang \& Wang}(2012)}]{Chuang2012}
  Chuang C.-H., Wang Y., 2012, MNRAS, 426, 226

\bibitem[{{Clifton et al.}(2012)}]{Clifton:2012}
  Clifton T., Ferreira P.G., Padilla A., Skordis C., 2012, Phys. Rep., 513, 1

\bibitem[{{Conley et al.}(2011)}]{Conley2011}
  Conley A., et al., 2011, ApJS, 192, 1

\bibitem[{{Crittenden et al.}(2012)}]{Crittenden:2012}
  Crittenden R.G., Zhao G.-B., Pogosian L., Samushia L., Zhang X., 2012, JCAP, 02, 048

\bibitem[{{Daly \& Djorgovski}(2007)}]{Daly:2007}
  Daly R.A., Djorgovski S.G., 2007, Nucl. Phys. B Proc. Suppl., 173, 19

\bibitem[{{Daniel et al.}(2009)}]{Daniel2009}
  Daniel S.F., et al., 2009, PRD, 80, 023532

\bibitem[{{Davis et al.}(2011)}]{Davis2011}
  Davis M., et al., 2011, MNRAS, 413, 2906

\bibitem[{{de la Torre \& Guzzo}(2012)}]{delaTorre2012}
  de la Torre S., Guzzo L., 2012, MNRAS, 427, 327

\bibitem[{{Dutta \& Sorbo}(2008)}]{Dutta:2007}
  Dutta K., Sorbo L., 2008, Phys. Rev. D, 75, 063514

\bibitem[{{Dvali, Gabadadze \& Poratti}(2000)}]{Dvali2010}
  Dvali G., Gabadadze G., Poratti M., 2000, Physics Lett. B, 485, 208

\bibitem[{{Eisenstein et al.}(2005)}]{Eis05}
  Eisenstein D.J., et al., 2005, ApJ, 633, 560

\bibitem[{{Eisenstein et al.}(2011)}]{Eis11}
  Eisenstein D., et al., 2011, AJ, 142, 72

\bibitem[{{Fang, Hu \& Lewis}(2008)}]{Fang2008}
  Fang W., Hu W., Lewis A., 2008, PRD, 78, 087303

\bibitem[{{Frieman et al.}(1995)}]{Frieman:1995}
  Frieman J.A., Hill C.T., Stebbins A., Waga I., 1995, Phys. Rev. Lett., 75, 2077

\bibitem[{{Fukugita et al.}(1996)}]{Fukugita:1996}
  Fukugita M., et al., 1996, AJ, 111, 1748

\bibitem[{{Galli et al.}(2009)}]{Galli2009}
  Galli S., Melchiorri A., Smooth G.F., Zahn O., 2009, PRD, 80, 023508

\bibitem[{{Gonzalez-Morales et al.}(2011)}]{Gon11}
  Gonzalez-Morales A., Poltis R., Sherwin B., Verde L., [arxiv:1106.5052] 

\bibitem[{{Gunn et al.}(1998)}]{Gunn:1998}
  Gunn J.E., et al., 1998, AJ, 116, 3040

\bibitem[{{Gunn et al.}(2006)}]{Gunn:2006}
  Gunn J.E., et al., 2006, AJ, 131, 2332

\bibitem[{{Guzzo et al.}(2007)}]{Guz08}
  Guzzo L., et al., 2008, Nature, 451, 541

\bibitem[{{Hamilton}(1998)}]{HamiltonReview}
  Hamilton A.J.S., ``Linear redshift distortions: A review'', in ``The
  Evolving Universe'', ed.~D.~Hamilton, pp.~185-275 (Kluwer Academic,
  1998) [astro-ph/9708102]

\bibitem[{{Hudson \& Turnbull}(2012)}]{Hudson2012}
  Hudson M.J., Turnbull S.J., 2012, ApJL, 715, 30

\bibitem[{{Ishak, Upadhye \& Spergel}(2006)}]{Ishak:2006}
  Ishak M., Upadhye A., Spergel D.N., 2006, PRD, 74, 043513

\bibitem[{{Ishak \& Dossett}(2009)}]{Ishak2009}
  Ishak M., Dossett J., 2009, PRD, 80, 043004

\bibitem[{{Jimenez, Talavera \& Verde}(2012)}]{Jimenez2011}
  Jimenez R., Talavera P., Verde L., 2012, Int. J. Modern Phys. A, 27, 1250174

\bibitem[{{Jimenez et al.}(2012)}]{Jimenez2012}
  Jimenez R., et al., 2012, JCAP, 03, 014

\bibitem[{{Kaiser}(1987)}]{Kai87}
  Kaiser N., 1987, \mnras, 227, 1

\bibitem[{{Kawasaki, Moroi \& Takahashi}(2001)}]{Kawasaki:2001}
  Kawasaki M., Moroi T., Takahashi T., 2001, Phys. Rev. D, 64, 083009

\bibitem[{{Komatsu et al.}(2011)}]{Komatsu2011}
  Komatsu E., et al., 2011, ApJS, 192, 18

\bibitem[{{Lampeitl et al.}(2010)}]{Lampeitl2010}
  Lampeitl H., et al., 2010, ApJ, 722, 566

\bibitem[{{Larson et al.}(2011)}]{WMAP7}
  Larson D., et al., 2011, ApJS, 192, 16

\bibitem[{{Lazkoz, Salzano \& Sendra}(2012)}]{lazkoz2012}
  Lazkoz R., Salzano V., Sendra I., 2012, EJPC, 72, 2130

\bibitem[{{Lewis, Challinor \& Lasenby}(2000)}]{Lewis2000}
  Lewis A., Challinor A., Lasenby A., 2000, ApJ, 538, 473

\bibitem[{{Lewis \& Bridle}(2002)}]{Lewis2002}
  Lewis A., Bridle S., 2002, PRD, 66, 103511

\bibitem[{{Linder}(2003)}]{Linder2003}
  Linder E.V., 2003, PRL, 90, 091301

\bibitem[{{Linder}(2005)}]{Linder:2005}
  Linder E.V., 2005, Phys. Rev. D, 72, 043529

\bibitem[{{Linder \& Cahn}(2007)}]{Linder2007}
  Linder E.V., Cahn R.N., 2007, Astropart. Phys., 29, 336

\bibitem[{{Linder}(2008)}]{Linder:2008}
  Linder E.V., 2008, Gen. Rel. Grav., 2008, 40, 329

\bibitem[{{Manera et al.}(2012)}]{Manera:2012a}
  Manera M., et al., 2012, MNRAS, in press

\bibitem[{{Matsubara}(2011)}]{Matsubara2011}
  Matsubara T., 2011, Phys Rev D83, 083518

\bibitem[{{Montesano, Sanchez \& Phleps}(2011)}]{Montesano2011}
  Montesano F., Sanchez A.G., Phleps S., 2011, MNRAS, 421, 2656

\bibitem[{{Ng \& Wiltshire}(2001)}]{Ng:2001}
  Ng S.C.C., Wilthshire D.L., Phys. Rev. D, 2001, 63, 023503

\bibitem[{{Nuza et al.}(2012)}]{Nuza:2012}
  Nuza S.W., et al., 2012, preprint, [arxiv:1202.6057]

\bibitem[{{Okamura, Taruya \& Matsubara}(2011)}]{Okamura2011}
  Okamura T., Taruya A., Matsubara T., 2011, JCAP, 08, 012

\bibitem[{{Padmanabhan et al.}(2012)}]{Padmanabhan2012}
  Padmanabhan N., Xu X., Eisenstein D.J., Scalzo R., Cuesta A.J., Mehta K.T.,
  Kazin E., 2012, preprint, [arxiv:1202.0090]

\bibitem[{{Peebles}(1980)}]{Peebles1980}
  Peebles P.J., 1980, The Large-Scale Structure of the Universe. Princeton Univ. Press, Princeton, NJ

\bibitem[{{Peebles \& Ratra}(1988)}]{Peebles:1988}
  Peebles P.J., Ratra B., 1988, ApJL, 325, 17

\bibitem[{{Peebles \& Ratra}(2003)}]{Peebles2003}
  Peebles P.J., Ratra B., 2003, Rev. Mod. Phys., 75, 559

\bibitem[{{Percival et al.}(2004)}]{Per04}
  Percival W.J., et al., 2004, MNRAS, 353, 1201

\bibitem[{{Percival et al.}(2010)}]{Per10}
  Percival W.J., et al., 2010, MNRAS, 401, 2148

\bibitem[{{Rapetti et al.}(2012)}]{Rapetti2012}
  Rapetti D., et al., 2012, preprint, [arxiv:1205.4679]

\bibitem[{{Ratra \& Peebles}(1988)}]{ratra88}
  Ratra B.V., Peebles P.J.E., 1988, PRD, 37, 3406

\bibitem[{{Reid et al.}(2010)}]{Rei10}
  Reid B.A., et al., 2010, MNRAS, 404, 60

\bibitem[{{Reid \& White}(2011)}]{ReiWhi11}
  Reid B.A., White M., 2011, MNRAS, 417, 1913

\bibitem[{{Reid et al.}(2012)}]{Reid12}
  Reid B.A., et al., 2012, MNRAS, 426, 2719

\bibitem[{{Riess et al.}(2011)}]{riess:2011}
  Riess, A., et al, 2011, Astrophys. J., 730, 119

\bibitem[{{Ross et al.}(2012)}]{Ross:2011}
  Ross A., et al., 2012, MNRAS, 424, 564

\bibitem[{{Sahni, Shafieloo \& Starobinsky}(2008)}]{Sahni:2008}
  Sahni V., Shafieloo A., Starobinsky A.A., 2008, PRD, 78, 3502

\bibitem[{{Samushia}(2009)}]{samushiathesis}
  Samushia L., 2009, PhD thesis, Kansas State University

\bibitem[{{Samushia et al.}(2011)}]{Samushia11}
  Samushia L., et al., 2011, MNRAS, 410, 1993

\bibitem[{{Samushia et al.}(2012)}]{SamPerRac11}
  Samushia, L., Pericval, W.J., Raccanelli, A., 2012, MNRAS, 420, 2102 

\bibitem[{{Sanchez et al.}(2012)}]{Sanchez:2012}
  Sanchez A.G., et al., 2012, MNRAS, 425, 415

\bibitem[{{Schlegel, White \& Eisenstein}(2009)}]{BOSS}
  Schlegel D., White M., Eisenstein D., 2009, The Astronomy and
  Astrophysics Decadal Survey, Science White Papers \#314 [arxiv:0902.4680]

\bibitem[{{Shafieloo et al.}(2006)}]{Shafieloo:2006}
  Shafieloo A., Alam 

\bibitem[{{Shafieloo \& Clarkson}(2010)}]{Shafieloo:2010}
  Shafieloo A., Clarkson C., 2010, PRD, 81, 3512

\bibitem[{{Shapiro et al.}(2010)}]{Shapiro:2010}
  Shapiro C., et al., 2010, PRD, 82, 043520

\bibitem[{{Song et al.}(2011a)}]{Song2011}
  Song Y.-S., Zhao G.-b., Bacon D., Koyama K., Nichol R.C., Pogosian L., 2011a, Phys. Rev. D, 84, 083523

\bibitem[{{Song et al.}(2011b)}]{Song:2011}
  Song Y.-S., Sabiu C.G., Kayo I., Nichol R.C., 2011b, JCAP, 1105, 020

\bibitem[{{Starobinsky}(2007)}]{Starobinsky2007}
  Starobinsky A.A., 2007, JTEPL, 86, 157

\bibitem[{{Stompor \& Efstathiou}(1999)}]{Stompor1999}
  Stompor R., Efstathiou G., 1999, MNRAS, 302, 735

\bibitem[{{Suzuki et al.}(2012)}]{Suzuki2011}
  Suzuki N., et al., 2012, ApJ, 746, 25

\bibitem[{{Taruya, Nishimichi \& Saito}(2010)}]{Taruya2010}
  Taruya A., Nishimichi T., Saito S., 2010, Phys Rev D82, 063522

\bibitem[{{Taruya, Saito \& Nishimichi}(2011)}]{Taruya2011}
  Taruya A., Saito S., Nishimichi T., 2011, Phys Rev D83, 103527

\bibitem[{{Tanner}(1996)}]{Tanner1996}
  Tanner M.A., 1996, ``Tools for statistical inference'', 3rd ed., Springer-Verlag, New York

\bibitem[{{Turnbull et al.}(2012)}]{Turnbull2012}
  Turnbull S.J., et al., 2012, MNRAS, 420, 447

\bibitem[{{Turner \& Huterer}(2007)}]{Turner:2007}
  Turner M.S., Huterer D., 2007, JPSJ, 11, 111015

\bibitem[{{Waga \& Frieman}(2000)}]{Waga:2000}
  Waga I., Frieman J.A., Phys. Rev. D, 2000, 62, 043521

\bibitem[{{Wang \& Steinhardt}(1998)}]{Wang:1998}
  Wang L., Steinhardt P.J., ApJ, 504, 483

\bibitem[{{Wang et al.}(2010)}]{Wang2010}
  Wang Y., et al., 2010, MNRAS, 409, 737

\bibitem[{{Weinberg et al.}(2012)}]{Weinberg2012}
  Weinberg D.H., Mortonson M.J., Eisenstein D.J., Hirata C., Riess A.G., Rozo
  E., 2012, preprint [arxiv:1201.2434]

\bibitem[{{White et al.}(2011)}]{Whi11}
  White M., et al., 2011, ApJ, 728, 126

\bibitem[{{York et al.}(2000)}]{York:2000}
  York D.J., et al., 2000, AJ, 120, 1579

\bibitem[{{Zaldarriaga \& Seljak}(1998)}]{Zaldarriaga1998}
  Zaldarriaga M., Seljak U., 1998, PRD, 58, 023003

\bibitem[{{Zhao et al.}(2009)}]{Zhao2009}
  Zhao G.-B., Pogosian L., Silvestri A., Zylberberg, J., 2009, PRL, 103, 241301

\bibitem[{{Zhao et al.}(2011)}]{zhao2011}
  Zhao G.-B., Hong L., Linder E.V., Koyama K., Bacon D.J., Zhang X., 2011, preprint, [arxiv:1109.1846]

\bibliographystyle{mnras}
\end{thebibliography}
\end{document}